# High-Temperature Superconductivity in Cerium Superhydrides

Wuhao Chen,[1] Dmitrii V. Semenok,[2] Xiaoli Huang,[1,*] Haiyun Shu[4], Xin Li,[1] Defang Duan,[1] Tian Cui,[3,1,*] and Artem R. Oganov[2]

[1] State Key Laboratory of Superhard Materials, College of Physics, Jilin University, Changchun 130012, China

[2] Skolkovo Institute of Science and Technology, Skolkovo Innovation Center, Bolshoy Boulevard 30, bld. 1 Moscow, Russia 121205

[3] School of Physical Science and Technology, Ningbo University, Ningbo 315211, China

[4] Center for High Pressure Science and Technology Advanced Research, Shanghai 201203, China

*Corresponding authors' e-mails: huangxiaoli@jlu.edu.cn, cuitian@nbu.edu.cn

**The discoveries of high-temperature superconductivity in $H_3S$ and $LaH_{10}$ have excited the search for superconductivity in compressed hydrides. In contrast to rapidly expanding theoretical studies, high-pressure experiments on hydride superconductors are expensive and technically challenging. Here we experimentally discover superconductivity in two new phases, $Fm\overline{3}m$-$CeH_{10}$ (SC-I phase) and $P6_3/mmc$-$CeH_9$ (SC-II phase) at pressures that are much lower (<100 GPa) than those needed to stabilize other polyhydride superconductors. Superconductivity was evidenced by a sharp drop of the electrical resistance to zero, and by the decrease of the critical temperature in deuterated samples and in an external magnetic field. SC-I has $T_c$=115 K at 95 GPa, showing expected decrease on further compression due to decrease of the electron-phonon coupling (EPC) coefficient λ (from 2.0 at 100 GPa to 0.8 at 200 GPa). SC-II has $T_c$ = 57 K at 88 GPa, rapidly increasing to a maximum $T_c$ ~100 K at 130 GPa, and then decreasing on further compression. This maximum of $T_c$ is due to a maximum of λ at the phase transition from $P6_3/mmc$-$CeH_9$ into a symmetry-broken modification $C2/c$-$CeH_9$. The pressure-temperature conditions of synthesis affect the actual hydrogen content, and the actual value of $T_c$. Anomalously low pressures of stability of cerium superhydrides make them appealing for studies of superhydrides and for designing new superhydrides with even lower pressures of stability.**

The search for high-temperature superconductivity is one of the most challenging tasks of condensed matter physics and materials science[1-5]. Achieving superconductivity at temperatures above the technologically important barrier of 77 K, the boiling point of liquid nitrogen ("high-temperature superconductivity"), has been one of the greatest triumphs of modern physics[6]. An appealing idea dating back to the early 1960s was that metallic hydrogen should be a high-temperature superconductor[7-10]. Due to the small mass of hydrogen, phonon frequencies in metallic hydrogen would be very high, of the order of several thousand kelvins[11], while covalent bonding will lead to strong electron-phonon coupling. However, extremely high pressures are needed for the transition of hydrogen from the molecular insulating phase into the metallic state[9,12,13]. For this reason, in recent years researchers have started to explore the possibility of inducing the superconducting state by adding small amounts of other elements to hydrogen[14-19], such that chemical precompression of hydrogen[20] results in strong reduction of the metallization pressure while maintaining high $T_c$.

Several hydrogen-rich compounds with different values of $T_c$ have been synthesized in experiment[21-27]. In 2014, an unusual high-pressure compound $H_3S$ with superconductivity at 191–204 K has been predicted[28] and later experimentally obtained[24,29], starting a new era in studies of superconductivity. In 2019, this record of high-temperature superconductivity was broken, with $LaH_{10}$ experimentally proven to have nearly room-temperature superconductivity with $T_c$ of 250–260 K[21,22]. The high critical temperature is a result of strong interaction of electrons whose states belong to the wide band with high-frequency phonons (optical modes caused by the presence of light hydrogen ions)[30,31]. In a big step toward room-temperature superconductivity, the highest $T_c$ of 288 K at ~275 GPa, achieved by doping, has been reported recently in the C–S–H system[32]. Interestingly, all of these superconductors contain much more hydrogen than expected from atomic valences (e.g., $H_3S$ vs $H_2S$, $LaH_{10}$ vs $LaH_3$), and are stabilized at megabar pressures. While this almost excludes their practical use, studying such compounds may hint at compounds that can be room-temperature superconductors at mild or ambient pressures.

Recently, two groups successfully synthesized new cerium polyhydride $P6_3/mmc$-$CeH_9$ where each Ce atom is enclosed in a $H_{29}$ cage in the atomic hydrogen sublattice[4,5]. For this compound, superconductivity has not yet been studied experimentally, but there have been theoretical studies[5,33,34]. In this study, we experimentally discover superconductivity both in $CeH_9$ and in the newly synthesized $CeH_{10}$, and discuss the causes of quite different behavior of these superconductors in the context of both experiment and theory.

## Main paradigms

In our previous work[4,5], two synthetic paths to $P6_3/mmc$-CeH$_9$ were confirmed: both cold compression and high-temperature annealing of Ce in hydrogen (H$_2$). Here we used an improved technology, with ammonia borane (NH$_3$BH$_3$, or AB for shorthand) as the source of hydrogen, which is released on heating due to decomposition reaction NH$_3$BH$_3$→3H$_2$ + $c$-BN[35-37]. This approach has been successfully used in several recent studies[22,23,25,38,39].

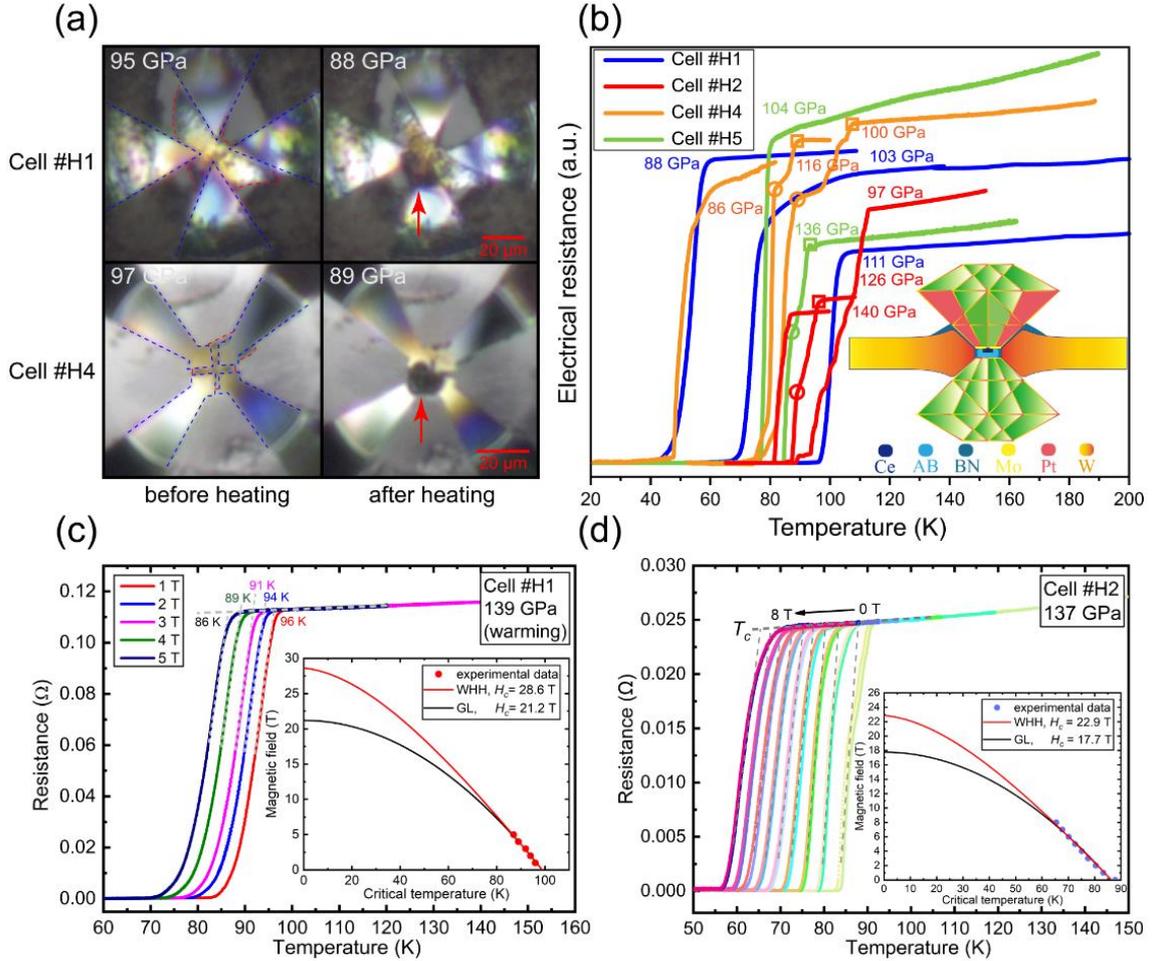

**Fig. 1.** Superconducting transitions determined by the electrical resistivity measurements in typical experimental diamond anvil cells. (a) Pressurized Ce/AB sample and four electrodes on the insulated gasket in the pressure cells #H1 and #H4 before and after laser heating. (b) Temperature dependence of the electrical resistance near the superconducting transitions (the warming cycle) in DACs #H1-5 at selected pressures (see details in Supporting Information). The scheme of the experimental assembly with a sample and electrodes prepared for the electrical resistance measurements is shown in the inset. (c, d) Temperature dependence of the electrical resistance in a magnetic field of 1–5 T at 139 GPa (cell #H1, warming cycle) and 0–8 T at 137 GPa (cell #H2, both cooling and warming cycles). $T_c$ was defined as the temperature at the onset of the transition. The upper critical magnetic field estimated using the Werthamer-Helfand-Hohenberg (WHH)[40] and Ginzburg-Landau (GL)[41] theories is shown in the inset.

**Superconductivity in Ce-H system.** We performed six experimental runs (cells #H1–H6) to investigate superconductivity in the Ce–H system at high pressures. The scheme of assembly used for the electrical measurements is shown in the inset in Fig. 1b. The details of the experimental methods and the parameters of diamond anvil cells (DACs) #H1–H6 are presented in Supporting Information. The Ce sample was placed inside the DAC with four deposited Mo electrodes and photographed in both reflected and transmitted light before and after laser heating at selected pressures (Fig. 1a). After laser heating, pressure decreased by about 7–8 GPa in both cell #H1 and #H4, whereas the volume of the sample increased at the heated part. The Raman vibron of $H_2$ was detected, indicating a local excess of hydrogen necessary to form cerium polyhydrides (Supporting Information Fig. S4b and Fig. S15 Inset). The obtained values of $T_c$ are quite close between cooling and warming cycles except cell #H1, where we faced a problem with the temperature controller. Quite a sharp ($\Delta T_c$=5-10 K) superconducting transition is observed in most of the $R$–$T$ curves obtained at various pressures in different experimental cells (Fig. 1b). In cell #H1, pressure decreased from 95 GPa to 88 GPa after laser heating and the electrical resistance dropped sharply to zero at 49 K (cooling cycle, see Supporting Information Fig. S5). Upon warming at a controlled rate of about 1 K/min, the transition temperature increased to 57 K, also possibly because of an uncontrollable pressure change. Much higher values of the initial $T_c$, determined at the onset of the two-step superconducting transition, were observed for the samples in cells #H2 and #H3 heated at similar pressures (90–100 GPa), which may result from the presence of different CeH$_x$ phases. We tentatively propose two obvious critical transition temperatures $T_{c1}$ and $T_{c2}$ which are discussed below. Upon further compression of the sample in cell #H2, the superconducting transition width decreases, reflecting different pressure dependences of $T_c$ of the phases synthesized in cell #H2. Noteworthy, the transition is reversible during decompression (Supporting Information Fig. S9 Inset).

We compared the $R$-$T$ curves of the samples in cells #H1 and #H2 with applied external magnetic field at the pressure of ~140 GPa (Fig. 1c, d). In cell #H1, $T_c$ decreases from 96 to 86 K as the applied magnetic field is raised from 1 to 5 T (Fig. 1c), yielding the slope of the critical field d$H_c$/d$T_c$ = -0.4 T/K. The sharp drop of resistance and the shift of the transition to lower temperatures in the magnetic field give strong evidence that synthesized polyhydrides do become superconducting at pressures above 88 GPa. The applied magnetic fields are insufficient to directly determine the upper critical magnetic field ($H_{c2}$) at 0 K, therefore we obtained $H_{C2}(0)$ by extrapolation using Ginzburg–Landau

(GL)[41,42] and Werthamer–Helfand–Hohenberg (WHH)[40] formulae, and obtained of 21.2 and 28.6 T, respectively. In cell #H2, the superconducting nature of the transitions was verified similarly by their dependence on the external magnetic field in the range of 0–8 T. The applied magnetic field of 8 T lowers $T_c$, defined as the temperature at the onset of the superconducting transition, by 23 K. Extrapolations based on GL and WHH expressions yield $H_{C2}(0) \sim 17.7$ and 22.9 T, respectively, which agrees well with the predicted value of ~22 T (linear interpolation between 120 and 150 GPa, Supporting Table S3). To increase $T_c$, we heated cell #H3 at 100 GPa for several times and reheated cells #H4 and #H5 at 100 and 136 GPa, respectively. After reheating, steps in the $R$–$T$ plot appeared and $T_c$ increased, whereas cell #H3 broke (Supporting Information Fig. S12).

**Crystal structure of superconducting phases.** To determine the crystal structure of obtained cerium hydrides, we analyzed their synchrotron X-ray diffraction (XRD) patterns. Although XRD patterns can be used to extract the exact positions of heavy Ce atoms, the positions of hydrogen atoms cannot be determined using this method because of a very low atomic X-ray scattering factor of hydrogen. Theoretical calculations help us to determine the crystal structure and stoichiometry through predictions of stable phases and pressure–volume relations. Besides $P6_3/mmc$-CeH$_9$, two cubic phases $F\bar{4}3m$-CeH$_9$[5,33] and $Fm\bar{3}m$-CeH$_{10}$[5,33,34,43] were also predicted to be stable above 90 and 170 GPa, respectively, while not being reported in previous experiments. In this study, the experimental XRD patterns show the presence of $I4/mmm$-CeH$_4$ and $P6_3/mmc$-CeH$_9$ in all the cells #H1–H6 (Fig. 2a). At 172 GPa for cell #H6, the highest pressure we studied, a set of diffraction peaks from cubic phase strikingly appeared. Comparing the unit cell volume of cerium hydrides with theoretical calculations, we confirmed the presence of the new cubic phase $Fm\bar{3}m$-CeH$_{10}$. Appearance of this phase at much lower pressures than predicted could be due to quantum motion of H atoms in the anharmonic potential[31]. Besides, delocalized nature of $f$ electrons may also contribute to the lower experimental stable pressure of CeH$_{10}$[44]. Lattice parameters and unit cell volumes of these phases are shown in Supporting Table S2. The equations of state of these polyhydrides obtained by XRD are in agreement with theoretical calculations (Fig. 2b), which further confirms the topology of the Ce sublattice and stoichiometry of the synthesized phases. Thus, the presence of two high-temperature superconducting phases, $P6_3/mmc$-CeH$_9$ and $Fm\bar{3}m$-CeH$_{10}$, can explain the two-step resistance drops during the superconducting transition observed in cells #H2, #H4, and #H5.

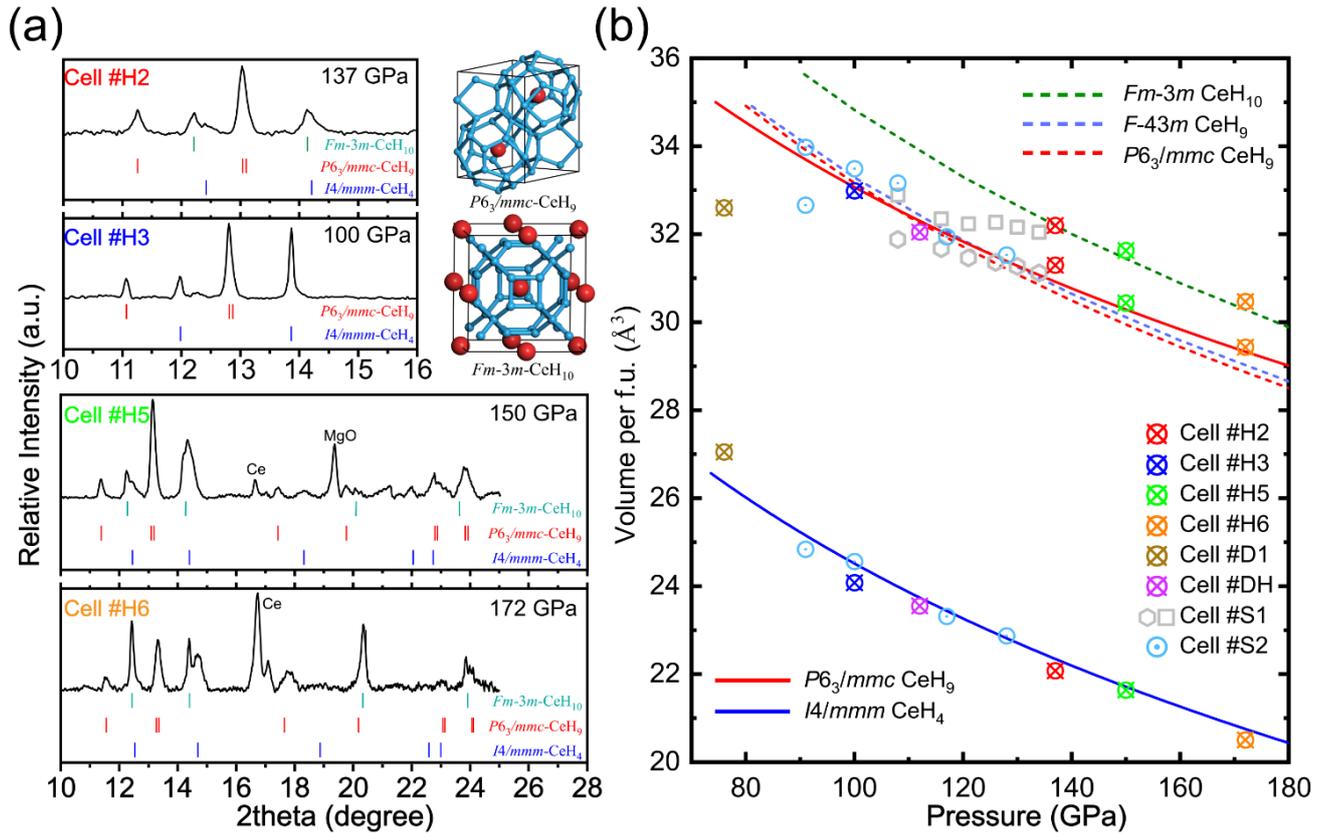

**Fig. 2.** XRD patterns and *P–V* equations of state of the Ce–H phases at different pressures. (a) Typical XRD patterns of the synthesized phases in the electrical cells #H2, #H3, #H5, and #H6, and the incident wavelength is 0.6199 Å. (b) Pressure dependence of the unit cell volume per formula unit. Experimental data in this study are represented by different point symbols. Dashed and solid lines indicate the calculated *P–V* data and fitted experimental data from Ref.[4], respectively. More data are plotted in Fig. S1.

**The isotope effect.** According to Bardeen–Cooper–Schrieffer (BCS) theory, conventional superconductors should exhibit the isotope effect: Decrease in the critical temperature with increasing atomic mass (which affects phonon frequencies). Thus, we prepared fully deuterated samples and studied them. Considering that this substitution may influence the stability of Ce–H compounds, as has been reported for LaH$_{10}$[21], we decided to perform structural investigation first and studied three diamond anvil cells #S1–S3. Cell #S1 was loaded with a Ce particle and deuterium gas (see Supporting Information for cell #S1). In this cell *Fm$\overline{3}$m*-CeD$_3$ was synthesized at 12 GPa right after loading D$_2$. When pressure was further increased to 58 GPa, tetragonal *I4/mmm*-CeD$_4$ formed. In contrast to the formation of *P6$_3$/mmc*-CeH$_9$ at around 100 GPa, CeD$_4$ remains stable up to 140 GPa even it was surrounded with a sufficient amount of deuterium (Fig. S23). With additional laser heating of the sample at 140 GPa, we synthesized both *P6$_3$/mmc*-CeD$_9$ and *Fm$\overline{3}$m*-CeD$_{10}$ (Fig. S22). As shown in Fig. 2b and Fig. S1, reducing pressure in the cell #S1 leads to a series of successive transformations:

at 116 GPa $Fm\bar{3}m$-CeD$_{10}$ lost one deuterium and turned into $F\bar{4}3m$-CeD$_9$ which finally decomposes at about 90 GPa (gray squares). At the same time, in the hexagonal phase the hydrogen desorption continues below 100 GPa (gray hexagons).

In cells #S2 and S3, Ce particles were sandwiched in the fully deuterated ammonia borane (d-AB). Laser heating of the sample in cell #S2 at 91 GPa resulted in a mixture of $\beta$-$Pm\bar{3}n$-CeD$_3$, $I4/mmm$-CeD$_4$, and the target compound $P6_3/mmc$-CeD$_9$. The evolution of the XRD pattern of the main detected phase ($hcp$) at different pressures is shown in Fig. 3a. We found (100) and (101, 002) reflections splitting in the pattern of $hcp$ phase to disappear at 117 GPa. We believe the peak splitting is observed because the deuterium content in the synthesized $hcp$ phases depends on pressure, heating conditions and the amount of D$_2$ released from d-AB. The same behavior was observed for $hcp$-I and $hcp$-II modifications of LaH$_{10}$ (Ref.[45], Figure 1a). Also, we verified that $P6_3/mmc$-CeD$_9$ can be synthesized from Ce and d-AB with laser heating.

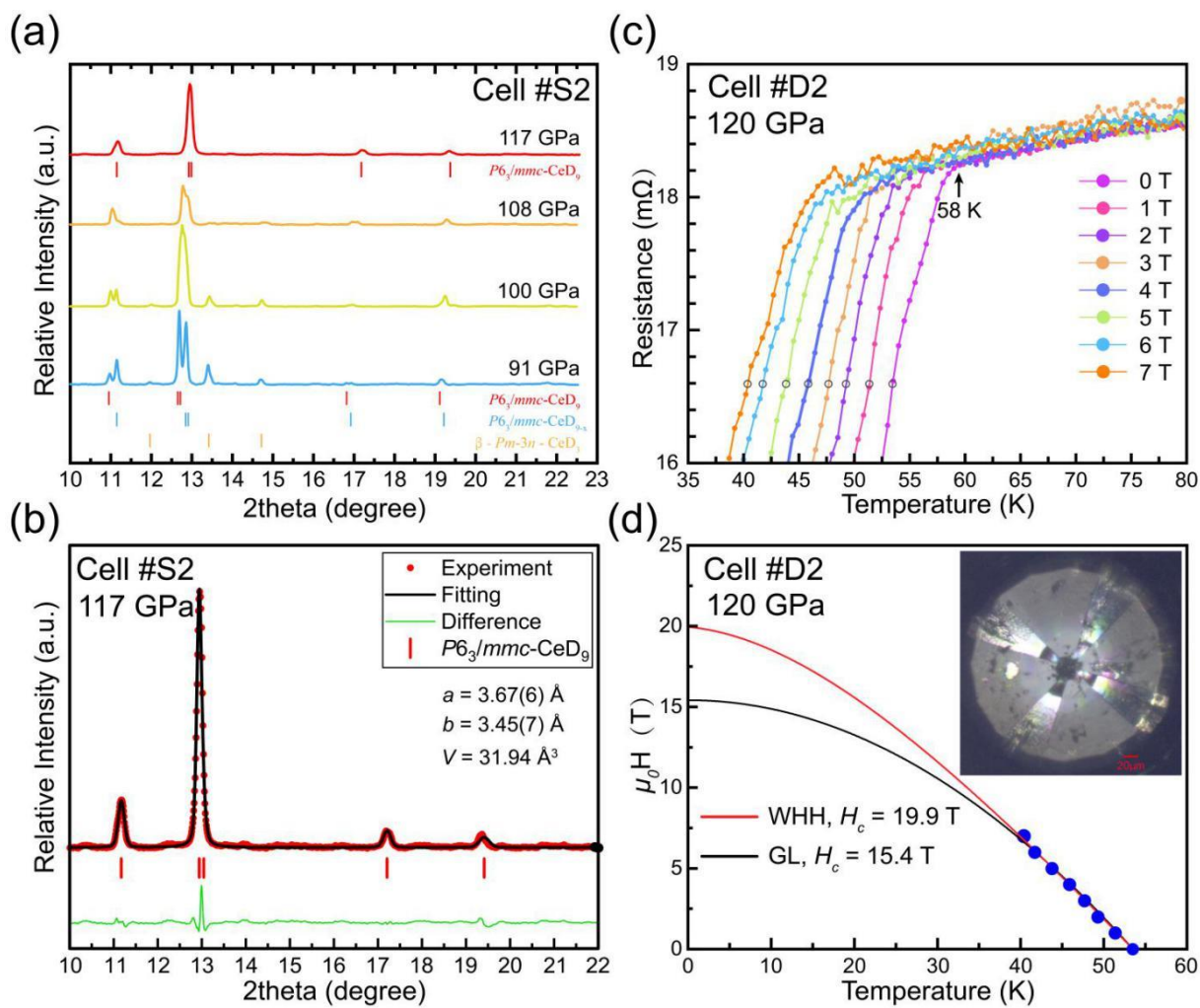

**Fig. 3.** XRD patterns and a series of superconducting transitions in external magnetic field for the synthesized Ce–D phases. (a) XRD patterns (wavelength $\lambda = 0.6199$ Å) of the obtained cerium deuterides at selected pressures in cell #S2. (b) Le Bail refinement of XRD pattern in cell #S2 at 117 GPa and $P6_3/mmc$-CeD$_9$ phase is identified. (c) Temperature dependence of the electrical resistance of the sample in cell #D2 in an external magnetic field of 0–7 T at 120 GPa. Circles represent the selected data for extrapolating $H_{c2}(0)$. (d) Upper critical magnetic field was extrapolated using the WHH[40] and GL[41,42] models for the sample in cell #D2 at 120 GPa. Inset is the photo of sample chamber.

The change in $T_c$ resulting from the substitution of hydrogen with deuterium provides direct evidence of the superconducting pairing mechanism. We measured the isotope effect using the electrical cells #D1–#D3. For the sample in cell #D1, $T_c = 25$ K was detected at 76 GPa, and its XRD pattern revealed the composition is a mixture of $\beta$-$Pm\bar{3}n$-CeD$_3$ and $P6_3mc$-CeD$_8$ (Fig. S26). The sample in cell #D3 shows a peculiarity in the $R$–$T$ curve around 40 K at 100 GPa (Fig. S28). For the sample in cell #D2, the onset of the superconducting transition was observed at 35 K, 100 GPa (Fig. S27) and 58 K, 120 GPa (Fig. 3c). The upper critical magnetic field estimated using the WHH and GL models reaches 19.9 T and 15.4 T, respectively (Fig. 3d), which is in close agreement with the calculated value (~17 T at 120 GPa). We have determined the isotope coefficient $\alpha$ (defined as $T_c(CeD_x) = T_c(CeH_x) \times m^{-\alpha}$), where $m$ is D/H mass ratio. With $T_{c1} \approx 82$ K for $P6_3/mmc$-CeH$_9$ in cell #H4 or #H5, and $T_{c2} \approx 58$ K in cell #D2, the estimated isotope coefficient at 120 GPa is equal to 0.49, close to theoretically predicted values (see Supplementary Table S3). Thus, BCS theory explains well the experimental upper critical magnetic fields for $P6_3/mmc$-CeH$_9$ and CeD$_9$, and gives correct value for the isotope coefficient.

**The pressure dependences of $T_c$.** The earliest theoretical results for $P6_3/mmc$-CeH$_9$ have been reported in 2017 by Peng et al.[34], who found $T_c = 56$ K at 100 GPa. However, Salke et al.[5] predicted that $P6_3/mmc$-CeH$_9$, with calculated $T_c = 105$-117 K at 200 GPa should be unstable at pressures below 120 GPa, where monoclinic $C2/c$-CeH$_9$ (which is a distortion of $P6_3/mmc$-CeH$_9$) is stable, with much lower $T_c = 63$-75 K at 100 GPa. Simultaneously, $P6_3mc$-CeH$_8$ was predicted stable at 55-95 GPa. For $Fm\bar{3}m$-CeH$_{10}$, another work predicted maximum $T_c = 168$ K at 94 GPa [33]. Thus, from theoretical calculations one can expect that below 100 GPa, $Fm\bar{3}m$-CeH$_{10}$ will have a higher $T_c$ compared with $P6_3/mmc$-CeH$_9$. Figs. 4a and 4c summarize the experimental pressure dependences of $T_c$ of $Fm\bar{3}m$-CeH$_{10}$ (SC-I) and $P6_3/mmc$-CeH$_9$ (SC-II). For phase SC-I, $T_c$ reaches the maximum of 115 K at the lowest pressure, 95 GPa, then decreases linearly as pressure increases. The decompression $T_c$–$P$ data for the sample in cell #H2 further confirm this tendency. Phase SC-II shows a different trend, displaying a dome-like $T_c(P)$ dependence (Fig. 4c) observed earlier for $H_3S$[46] and $LaH_{10}$[21,45], with a maximum at 130 GPa.

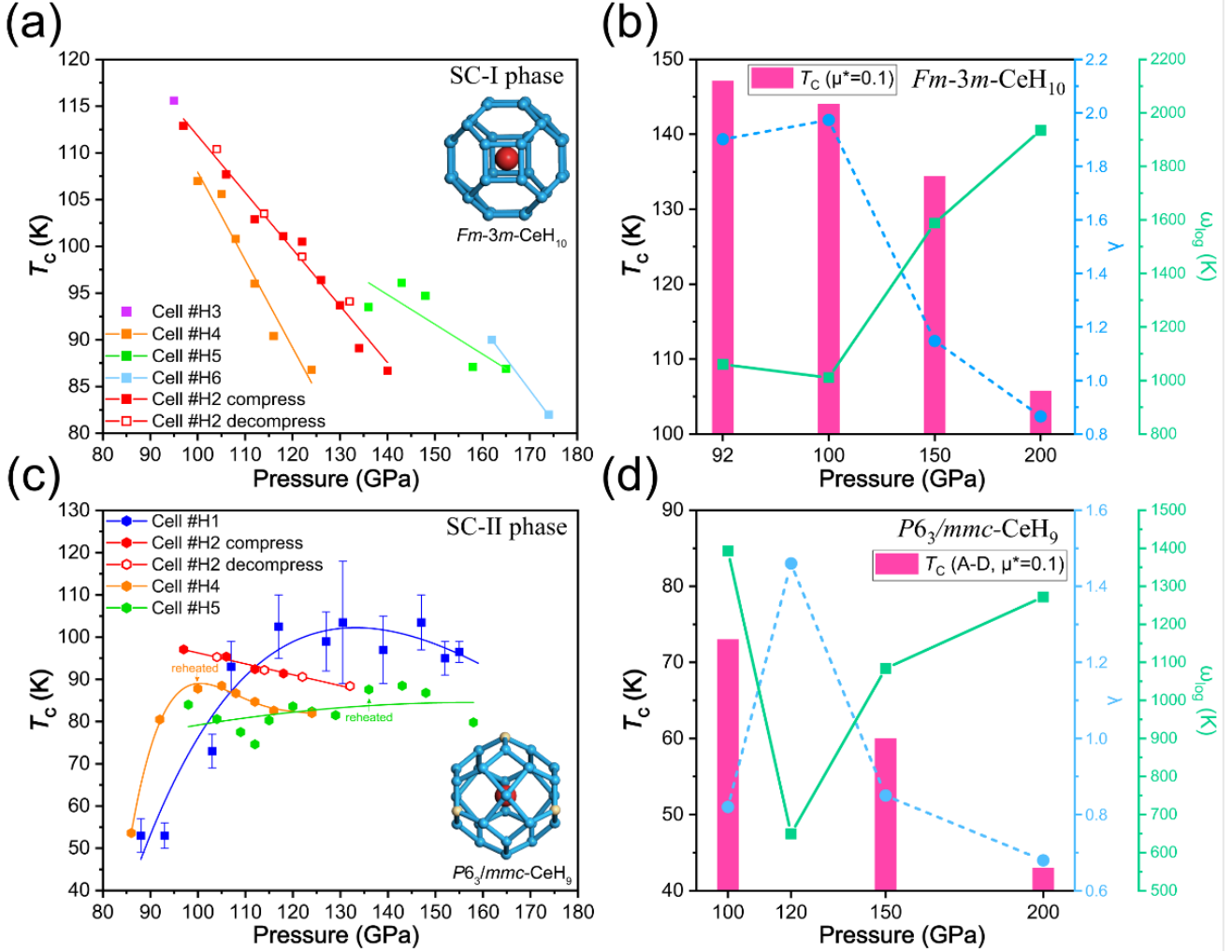

**Fig. 4.** Superconducting parameters of $Fm\overline{3}m$-$CeH_{10}$ and $P6_3/mmc$-$CeH_9$. (a, c) Experimental $T_c$ as a function of pressure for phases SC-I and SC-II. Insets show the hydrogen cages and yellow atoms represents the lost hydrogen from $P6_3/mmc$-$CeH_9$ to $P6_3mc$-$CeH_8$. (b, d) Calculated logarithmic averaged frequency $\omega_{log}$, EPC coefficient $\lambda$, and $T_c$ at various pressures for $Fm\overline{3}m$-$CeH_{10}$ and $P6_3/mmc$-$CeH_9$. The theoretical data for $Fm\overline{3}m$-$CeH_{10}$ are from Ref. [33]

The spread of critical temperatures obtained in different experimental DACs requires some discussion. Superconductivity is very sensitive to stoichiometry and structural distortions, and deviation from the ideal stoichiometric Ce: H ratio and consequent variation of the degree of distortion are the likeliest causes of the variation of results obtained from different cells. We have previously observed a continuous increase of the unit cell volume during compression from 80 to 103 GPa without laser heating[4] that represents the gradually increasing of hydrogen stoichiometry to the ideal ratio 1:9. In contrast, Salke et al.[5] reported direct formation of CeH$_9$ after heating to over 1700 K at 80 GPa. This indicates the competing between $P6_3mc$-CeH$_8$ and $C2/c$-CeH$_9$, depending on the heating temperature, thus CeH$_{8-9}$ was produced. In another aspect, instead of loading hydrogen, AB was used in this study that we cannot ensure the released H$_2$ is sufficient for each cell. Heating procedure will also affect the distortion degree of $C2/c$-CeH$_9$, while it is indistinguishable in X-ray diffraction (the same situation as in H$_3$S[46] and LaH$_{10}$[45]).

As shown in Fig. 4d and Table S3, *ab initio* calculations for $P6_3/mmc$-CeH$_9$ demonstrate that in the pressure range from 100 to 120 GPa, the EPC coefficient λ increases from 0.82 to 1.46, and $T_c$ increases from 73 to 82 K (μ* = 0.1) in agreement with our experimental results. For pressures above 120 GPa, stabilization of $P6_3/mmc$-CeH$_9$ leads to an increase in the logarithmic frequency ω$_{log}$ and the Debye temperature with a simultaneous decrease in the critical temperature and the EPC coefficient. The same situation occurs for $Fm\bar{3}m$-CeH$_{10}$ in the pressure range of 100–200 GPa (Fig. 4b), as λ drops from 2.0 to 0.8 and $T_c$ decreases monotonically, in close agreement with the experimental tendency (Fig. 4a). This behavior, caused by weakening of the electron–phonon coupling, is common in conventional phonon-mediated superconductors such as the two-band superconductor MgB$_2$[47].

**Conclusions**

Synthesis and comprehensive study of superconductivity of superhydrides and superdeutrides ( $Fm\bar{3}m$ -CeH$_{10}$, $P6_3/mmc$-CeH$_9$, $Fm\bar{3}m$ -CeD$_{10}$, $F\bar{4}3m$ -CeD$_9$, $P6_3/mmc$-CeD$_9$, $P6_3mc$-CeD$_8$, $P6_3mc$-CeD$_6$ and $I4/mmm$-CeD$_4$) showed that cerium hydrides are remarkable compounds. Stable and displaying high-temperature superconductivity at lower pressures than any other superhydrides, they serve as an ideal starting point to further study the mechanism of superconductivity in these fascinating compounds, and design other superconductors, stable at even lower pressures. Soft-mode-driven phase transition in CeH$_9$ is responsible for a maximum in $T_c$, and this can be used for engineering of higher-temperature superconductors. Away from the phase transition, $T_c$ of both CeH$_9$ and CeH$_{10}$ decreases with pressure, leading one to hope that search for superconductors with lower-pressure stability can lead to increased $T_c$.

## Acknowledgments

In situ angle dispersive XRD were performed at 4W2 HP-Station, Beijing Synchrotron Radiation Facility (BSRF), and 15U1 beamline, Shanghai Synchrotron Radiation Facility (SSRF). This work was supported by the National Key R&D Program of China (No. 2018YFA0305900), National Natural Science Foundation of China (Nos. 51572108, 11974133, 11804113, 51720105007), and Program for Changjiang Scholars, Innovative Research Team in University (No. IRT_15R23). D.V.S. was funded by the Russian Foundation for Basic Research (project 20-32-90099). A.R.O is supported by the Ministry of Science and Higher Education (grant 2711.2020.2 to leading scientific schools). We thank Igor Grishin (Skoltech) for proofreading of the manuscript.

## References

1    Onnes, H. K. *Proc. K. Ned. Akad. Wet.* **13**, 1274 (1911).
2    Nagamatsu, J., Nakagawa, N., Muranaka, T., Zenitani, Y. & Akimitsu, J. Superconductivity at 39 K in magnesium diboride. *Nature* **410**, 63-64 (2001).
3    Bednorz, J. G. & Müller, K. A. Possible highTc superconductivity in the Ba−La−Cu−O system. *Zeitschrift für Physik B Condensed Matter* **64**, 189-193 (1986).
4    Li, X. *et al.* Polyhydride CeH9 with an atomic-like hydrogen clathrate structure. *Nat. Commun.* **10**, 3461 (2019).
5    Salke, N. P. *et al.* Synthesis of clathrate cerium superhydride CeH9 at 80-100 GPa with atomic hydrogen sublattice. *Nat. Commun.* **10**, 4453 (2019).
6    Wu, M. K. *et al.* Superconductivity at 93 K in a new mixed-phase Y-Ba-Cu-O compound system at ambient pressure. *Phys. Rev. Lett.* **58**, 908-910 (1987).
7    Wigner, E. & Huntington, H. B. On the Possibility of a Metallic Modification of Hydrogen. **3**, 764-770 (1935).
8    McMahon, J. M., Morales, M. A., Pierleoni, C. & Ceperley, D. M. The properties of hydrogen and helium under extreme conditions. *Rev. Mod. Phys.* **84**, 1607-1653 (2012).
9    Azadi, S., Monserrat, B., Foulkes, W. M. C. & Needs, R. J. Dissociation of High-Pressure Solid Molecular Hydrogen: A Quantum Monte Carlo and Anharmonic Vibrational Study. *Phys. Rev. Lett.* **112**, 165501 (2014).
10   McMinis, J., Clay, R. C., Lee, D. & Morales, M. A. Molecular to Atomic Phase Transition in Hydrogen under High Pressure. *Phys. Rev. Lett.* **114**, 105305 (2015).
11   Ashcroft, N. W. Metallic Hydrogen: A High-Temperature Superconductor? *Phys. Rev. Lett.* **21**, 1748-1749 (1968).
12   Cudazzo, P. *et al.* Ab Initio Description of High-Temperature Superconductivity in Dense Molecular Hydrogen. *Phys. Rev. Lett.* **100**, 257001 (2008).
13   Monserrat, B. *et al.* Structure and Metallicity of Phase V of Hydrogen. *Phys. Rev. Lett.* **120**, 255701 (2018).

14    Pickard, C. J., Errea, I. & Eremets, M. I. Superconducting Hydrides Under Pressure. *Annu. Rev. Condens. Matter Phys.* **11**, 57-76 (2020).

15    Semenok, D. V., Kruglov, I. A. & Kvashnin, A. G. On Distribution of Superconductivity in Metal Hydrides. *Curr. Opin. Solid State Mater. Sci.* **24**, 100808 (2020).

16    Duan, D. *et al.* Structure and superconductivity of hydrides at high pressures. *Natl. Sci. Rev.* **4**, 121-135 (2016).

17    Hutcheon, M. J., Shipley, A. M. & Needs, R. J. Predicting novel superconducting hydrides using machine learning approaches. *Phys. Rev. B* **101**, 144505 (2020).

18    Zurek, E. & Bi, T. High-temperature superconductivity in alkaline and rare earth polyhydrides at high pressure: A theoretical perspective. *J. Chem. Phys.* **150**, 050901 (2019).

19    Flores-Livas, J. A. *et al.* A perspective on conventional high-temperature superconductors at high pressure: Methods and materials. *Phys. Rep.* **856**, 1-78 (2020).

20    Ashcroft, N. W. Hydrogen dominant metallic alloys: High temperature superconductors? *Phys. Rev. Lett.* **92**, 187002 (2004).

21    Drozdov, A. P. *et al.* Superconductivity at 250 K in lanthanum hydride under high pressures. *Nature* **569**, 528-531 (2019).

22    Somayazulu, M. *et al.* Evidence for Superconductivity above 260 K in Lanthanum Superhydride at Megabar Pressures. *Phys. Rev. Lett.* **122**, 027001 (2019).

23    Semenok, D. V. *et al.* Superconductivity at 161 K in thorium hydride ThH10: Synthesis and properties. *Mater. Today* **33**, 36-44 (2020).

24    Drozdov, A. P., Eremets, M. I., Troyan, I. A., Ksenofontov, V. & Shylin, S. I. Conventional superconductivity at 203 kelvin at high pressures in the sulfur hydride system. *Nature* **525**, 73 (2015).

25    Troyan, I. A. *et al.* Anomalous high-temperature superconductivity in YH6. *Preprint at https://arxiv.org/abs/1908.01534* (2020).

26    Kong, P. P. *et al.* Superconductivity up to 243 K in yttrium hydrides under high pressure. *Preprint at https://arxiv.org/abs/1909.10482* (2019).

27    Drozdov, A. P., Eremets, M. I. & Troyan, I. A. Superconductivity above 100 K in PH3 at high pressures. *Preprint at https://arxiv.org/abs/1508.06224* (2015).

28    Duan, D. F. *et al.* Pressure-induced metallization of dense (H2S)(2)H-2 with high-T-c superconductivity. *Sci. Rep.* **4**, 6968 (2014).

29    Huang, X. L. *et al.* High-temperature superconductivity in sulfur hydride evidenced by alternating-current magnetic susceptibility. *Natl. Sci. Rev.* **6**, 713-718 (2019).

30    Liu, H. Y., Naumov, I. I., Hoffmann, R., Ashcroft, N. W. & Hemley, R. J. Potential high-T-c superconducting lanthanum and yttrium hydrides at high pressure. *Proc. Natl. Acad. Sci. U. S. A.* **114**, 6990-6995 (2017).

31    Errea, I. *et al.* Quantum crystal structure in the 250-kelvin superconducting lanthanum hydride. *Nature* **578**, 66-69 (2020).

32    Snider, E. *et al.* Room-temperature superconductivity in a carbonaceous sulfur hydride. *Nature* **586**, 373-377 (2020).

33    Li, B. *et al.* Predicted high-temperature superconductivity in cerium hydrides at high pressures. *J. Appl. Phys.* **126**, 235901 (2019).

34    Peng, F. *et al.* Hydrogen Clathrate Structures in Rare Earth Hydrides at High Pressures: Possible Route to Room-Temperature Superconductivity. *Phys. Rev. Lett.* **119**, 107001 (2017).

35    Kondrat'ev, Y. V., Butlak, A. V., Kazakov, I. V. & Timoshkin, A. Y. Sublimation and thermal decomposition of ammonia borane: Competitive processes controlled by pressure. *Thermochim. Acta* **622**, 64-71 (2015).

36    Sun, Y., Chen, J., Drozd, V., Najiba, S. & Bollinger, C. Behavior of decomposed ammonia borane at high pressure. *J. Phys. Chem. Solids* **84**, 75-79 (2015).

37    Nylén, J. *et al.* Thermal decomposition of ammonia borane at high pressures. *J. Chem. Phys.* **131**, 104506 (2009).

38    Zhou, D. *et al.* High-Pressure Synthesis of Magnetic Neodymium Polyhydrides. *J. Am. Chem. Soc.* **142**, 2803-2811 (2020).

39    Zhou, D. *et al.* Superconducting praseodymium superhydrides. *Sci. Adv.* **6**, eaax6849 (2020).

40    Werthamer, N. R., Helfand, E. & Hohenberg, P. C. Temperature and Purity Dependence of the Superconducting Critical Field, Hc2. III. Electron Spin and Spin-Orbit Effects. *Phys. Rev.* **147**, 295-302 (1966).

41    Ginzburg, V. L. & Landau, L. D. On the Theory of superconductivity. *Zh.Eksp.Teor.Fiz.* **22**, 1064–1082 (1950).

42    Woollam, J. A., Somoano, R. B. & O'Connor, P. Positive Curvature of the Hc 2-versus-Tc Boundaries in Layered Superconductors. *Phys. Rev. Lett.* **32**, 712-714 (1974).

43    Tsuppayakorn-aek, P., Pinsook, U., Luo, W., Ahuja, R. & Bovornratanaraks, T. Superconductivity of superhydride CeH10 under high pressure. *Mater. Res. Express* **7**, 086001 (2020).

44    Jeon, H., Wang, C., Yi, S. & Cho, J.-H. Origin of enhanced chemical precompression in cerium hydride CeH9. *Sci. Rep.* **10**, 16878 (2020).

45    Sun, D. *et al.* High-temperature superconductivity on the verge of a structural instability in lanthanum superhydride. *Preprint at https://arxiv.org/abs/2010.00160* (2020).

46    Einaga, M. *et al.* Crystal structure of the superconducting phase of sulfur hydride. *Nat. Phys.* **12**, 835-838 (2016).

47    Buzea, C. & Yamashita, T. Review of the superconducting properties of MgB2. *Supercond. Sci. Technol.* **14**, R115-R146 (2001).

# SUPPORTING INFORMATION

## for

# High-Temperature Superconductivity in Cerium Superhydrides

Wuhao Chen,[1] Dmitrii V. Semenok,[2] Xiaoli Huang,[1,*] Haiyun Shu,[4] Xin Li,[1] Defang Duan,[1] Tian Cui,[3,1,*] and Artem R. Oganov[2]

[1] State Key Laboratory of Superhard Materials, College of Physics, Jilin University, Changchun 130012, China

[2] Skolkovo Institute of Science and Technology, Skolkovo Innovation Center, Bolshoy Boulevard 30, bld. 1 Moscow, Russia 121205

[3] School of Physical Science and Technology, Ningbo University, Ningbo, 315211, China

[4] Center for High Pressure Science and Technology Advanced Research, Shanghai 201203, China

*Corresponding authors' e-mails: huangxiaoli@jlu.edu.cn, cuitian@nbu.edu.cn

# Content

# I. Experimental and theoretical methods

**Sample preparation and electrical resistance measurements.** Cerium metal (purity 99.8%) was purchased from Alfa Aesar. The standard four-point probe electrical transport measurements were carried out using Mao type diamond anvil cells (DACs) made of nonmagnetic Cu–Be alloy in a multifunctional measurement system (1.5–300 K, JANIS Research Company Inc.; 0–9 T, Cryomagnetics Inc.). Diamond anvils with top flat of 50–100 μm beveled from 250–300 μm with an angle of 8.5° were used to generate ultrahigh pressure. Cubic boron nitride (c-BN) or magnesium oxide (MgO) powder mixed with an epoxy binder were used as an insulating layer between the tungsten gasket and platinum electrodes. Four Mo electrodes were sputtered onto the piston diamond and connected to the external wires using a combination of 25 μm thick Pt 'shoes' soldered onto brass holders. Ammonia borane $NH_3BH_3$ (AB), which we chose as the hydrogen source, was sublimed before use and loaded to diamond anvil cells inside the argon glovebox. When completely dehydrogenated, one mole of AB yields three moles of $H_2$ and insulating c-BN, the latter serving both as a solid pressure medium and support holding the hydride sample firmly against the electrical contacts. The Ce sample was positioned precisely on top of the electrodes inside the glovebox. The electrical contact between the sample and electrodes was confirmed upon compression. Pressure was determined using the dependence of the Raman shift of diamond[1].

Deuterated ammonium borane (d-AB) was synthesized from $NaBD_4$ (98 % D, SigmaAldrich) via reaction with ammonium formate $HCOONH_4$ in tetrahydrofuran followed by isotopic substitution (H→D) in $D_2O$ [2]. After removing of solvents and vacuum drying, the obtained $ND_3BD_3$ was analyzed by [1]H NMR and Raman spectroscopy. Deuterium content in the product was found to be 92 %. Partially substituted $NH_3BD_3$ and $ND_3BH_3$ were synthesized in a similar way for later use as a source of HD.

**Synchrotron X-ray diffraction measurements.** In situ high-pressure synchrotron X-ray diffraction (XRD) patterns were recorded on the BL15U1 synchrotron beamline at the Shanghai Synchrotron Research Facility (China) using a focused 5×12 μm monochromatic beam. Additional experiments with electrodes were carried out at the 4W2 High-Pressure Station of Beijing Synchrotron Radiation Facility (China) with the beam size of about 32×12 μm. Both facilities used an incident X-ray beam (20 keV, 0.6199 Å) and Mar165 two-dimensional charge-coupled device detector. The experimental XRD images were integrated and analyzed using Dioptas software package[3]. The full profile analysis of the diffraction patterns and calculation of the unit cell parameters were performed using Materials Studio[4] and Jana2006[5] programs with the Le Bail method[6].

**Theoretical calculations.** Calculations of superconducting $T_c$ of $P6_3/mmc$-CeH$_9$ were carried out using QUANTUM ESPRESSO (QE) package[7,8]. The phonon frequencies and electron–phonon coupling (EPC) coefficients were computed using the density functional perturbation theory[9], employing the plane-wave pseudopotential method with a cutoff energy of 80 Ry, and the Perdew–Burke–Ernzerhof or Goedecker–Hartwigsen–Hutter–Teter (for comparison) exchange–correlation functionals[9-11]. Within the optimized tetrahedron method[12], we calculated the electron–phonon coupling coefficients λ and the Eliashberg functions via sampling of the first Brillouin zone by $16 \times 16 \times 8$ $k$-points and $4 \times 4 \times 2$ $q$-points meshes. To evaluate the density of states (DOS) and electron–phonon linewidths, a much denser $24 \times 24 \times 16$ $k$-mesh was used.

The interpolation method[13] was used to calculate electron–phonon matrix elements of $P6_3/mmc$-CeH$_9$ at 150 and 200 GPa. Ultrasoft PBE pseudopotentials for Ce and H were used with a plane-wave basis set cutoff of 100 Ry. The $k$-space integration (electrons) was approximated by a summation over the $5 \times 5 \times 3$ uniform grid in reciprocal space for the self-consistent cycles; a much finer $20 \times 20 \times 12$ grid was used for evaluating the DOS and electron–phonon linewidths. Dynamical matrices and electron–phonon linewidths of CeH$_9$ were calculated on a uniform $5 \times 5 \times 3$ grid in $q$-space.

The critical temperature of superconducting transition was calculated using the Matsubara-type linearized Migdal–Eliashberg equations[14]:

$$\hbar \omega_j = \pi(2j+1)k_\text{B}T, \qquad j = 0, \pm 1, \pm 2, \ldots \tag{S1}$$

$$\lambda(\omega_i - \omega_j) = 2 \int\limits_0^\infty \frac{\omega \cdot \alpha^2 F(\omega)}{\omega^2 + (\omega_i - \omega_j)^2} d\omega \tag{S2}$$

$$\Delta(\omega = \omega_i, T) = \Delta_i(T)$$
$$= \pi k_\text{B}T \sum_j \frac{[\lambda(\omega_i - \omega_j) - \mu^*]}{\rho + |\hbar \omega_j + \pi k_\text{B}T \sum_k (sign\ \omega_k) \cdot \lambda(\omega_i - \omega_j)|} \cdot \Delta_j(T) \tag{S3}$$

where $T$ is the temperature in kelvins, $\mu^*$ is the Coloumb pseudopotential, $\omega$ is the frequency in Hz, $\rho(T)$ is a pair-breaking parameter, the function $\lambda(\omega_i - \omega_j)$ relates to an effective electron–electron interaction via the exchange of phonons. The transition temperature can be found by solving the equation $\rho(T_c) = 0$, where $\rho(T)$ is defined as $\max(\rho)$, provided that $\Delta(\omega)$ is not a zero function of $\omega$ at a fixed temperature.

These equations can be rewritten in a matrix form as[15]

$$\rho(T)\psi_m = \sum_{n=0}^{N} K_{mn}\psi_n \Leftrightarrow \rho(T)\begin{pmatrix} \psi_1 \\ ... \\ \psi_N \end{pmatrix} = \begin{pmatrix} K_{11} & ... & K_{1N} \\ ... & K_{ii} & ... \\ K_{N1} & ... & K_{NN} \end{pmatrix} \times \begin{pmatrix} \psi_1 \\ ... \\ \psi_N \end{pmatrix} \quad \text{(S4)}$$

where $\psi_n$ is related to $\Delta(\omega, T)$, and

$$K_{mn} = F(m-n) + F(m+n+1) - 2\mu^* - \delta_{mn}\left[2m+1+F(0)+2\sum_{l=1}^{m}F(l)\right] \quad \text{(S5)}$$

$$F(x) = F(x,T) = 2\int_0^{\omega \text{max}} \frac{\alpha^2 F(\omega)}{(\hbar\omega)^2 + (2\pi \cdot k_B T \cdot x)^2}\hbar\omega d\omega \quad \text{(S6)}$$

where $\delta_{nn} = 1$ and $\delta_{nm} = 0$ ($n \neq m$) are unit matrices. Now we can replace the equation $\rho(T_c) = 0$ with the vanishing of the maximum eigenvalue of the matrix $K_{nm}$:[$\rho = c(K_{nm}) = f(T)$, $f(T_c) = 0$]. More approximate estimates of $T_c$ were made using the Allen–Dynes (AD) formula[16].

To calculate the isotopic coefficient β, the McMillan formula was used:

$$\beta_{\text{McM}} = -\frac{d\ln T_c}{d\ln M} = \frac{1}{2}\left[1 - \frac{1.04(1+\lambda)(1+0.62\lambda)}{[\lambda - \mu^*(1+0.62\lambda)]^2}\mu^{*2}\right] \quad \text{(S7)}$$

The Sommerfeld constant was found as

$$\gamma = \frac{2}{3}\pi^2 k_B^2 N(E_F)(1+\lambda) \quad \text{(S8)}$$

and was applied to estimate the upper critical magnetic field and the superconductive gap in cerium superhydride using the well-known semiempirical equations of the BCS theory[17], Equations 4.1 and 5.11, which can be used for $T_c/\omega_{\text{log}} < 0.25$:

$$\frac{\gamma T_c^2}{B_{c2}^2(0)} = 0.168\left[1 - 12.2\left(\frac{T_c}{\omega_{\text{log}}}\right)^2\ln\left(\frac{\omega_{\text{log}}}{3T_c}\right)\right] \quad \text{(S9)}$$

$$\frac{2\Delta(0)}{k_B T_c} = 3.53\left[1 + 12.5\left(\frac{T_c}{\omega_{\text{log}}}\right)^2\ln\left(\frac{\omega_{\text{log}}}{2T_c}\right)\right] \quad \text{(S10)}$$

The lower critical magnetic field was calculated according to the Ginzburg–Landau theory[18]:

$$\frac{H_{c1}}{H_{c2}} = \frac{\ln k}{2\sqrt{2}k^2}, \qquad k = \lambda_L / \xi \qquad (S11)$$

where $\lambda_L$ is the London penetration depth:

$$\lambda_L = 1.0541 \times 10^{-5} \sqrt{\frac{m_e c^2}{4\pi n_e e^2}} \qquad (S12)$$

here $c$ is the speed of light, $e$ is the electron charge, $m_e$ is the electron mass, and $n_e$ is an effective concentration of charge carriers, evaluated from the average Fermi velocity ($V_F$) in the Fermi gas model:

$$n_e = \frac{1}{e\pi^2}\left(\frac{m_e V_F}{\hbar}\right)^3 \qquad (S13)$$

The coherence length $\xi$ was found as $\xi = \sqrt{\hbar/2e(\mu_0 H_{c2})}$ and was used to estimate the average Fermi velocity:

$$V_F = \frac{\pi \Delta(0)}{\hbar}\xi \qquad (S14)$$

It is possible to estimate the Debye temperature $\theta_D$ from the dependence of the electrical resistance $R(T)$ on the temperature, recorded in the interval between 300 K and the transition point, using fit of the experimental $R(T)$ by the Bloch–Grüneisen (BG) formula[19]:

$$R(T) = R_0 + A\left(\frac{T}{\theta_D}\right)^5 \int_0^{\frac{\theta_D}{T}} \frac{x^5}{(e^x - 1)(1 - e^{-x})} dx$$

where $A$, $\theta_D$, and $R_0$ were found using the least squares method. Recently, it has been demonstrated that when $R \to 0$ at $T < T_c$, the formula gives reasonable values of the Debye temperatures for $H_3S$ and $LaH_{10}$[20]. The analysis of the experimental data for $P6_3/mmc$-$CeH_9$ shows that $\theta_D$ and $\omega_{log}$ remain around 500–600 K at 130–160 GPa, which indicates the need of high $\lambda = 1.5$–1.8 to ensure that the observed critical temperature of superconductivity is above 100 K.

An independent evaluation of the Debye temperature and logarithmic frequency $\omega_{log}$ can be done using the elastic tensors of $CeH_9$, which were calculated using the stress–strain relations:

$$C_{ij} = \frac{\partial \sigma_i}{\partial \eta_j},$$ (S7)

where $\sigma_i$ and $\eta_j$ are the $i$th and $j$th components of the stress tensor, respectively.

The bulk ($B$) and shear ($G$) moduli and the Young's modulus ($E$) were calculated in GPa via the Voigt–Reuss–Hill averaging [21,22]. Using the obtained values of the elastic moduli, we calculated the velocities of the longitudinal and transverse acoustic waves:

$$v_{LA} = \sqrt{\frac{3B+4G}{3\rho}}, \ v_{TA} = \sqrt{\frac{G}{\rho}},$$ (S8)

where $B$, $G$ and $\rho$ are bulk modulus, shear modulus and a density of the compound, respectively.

The obtained values allow us to estimate the Debye temperature as [23]

$$\vartheta_D = \frac{h}{k_B} \left[ \frac{3n}{4\pi} \left( \frac{N_A \cdot \rho}{M} \right) \right]^{\frac{1}{3}} v_m$$ (S9)

where $h$, $k_B$, and $N_A$ are the Planck, Boltzmann, and Avogadro constants, $v_m$ is the average velocity of the acoustic waves calculated as

$$v_m = \left[ \frac{1}{3} \left( \frac{2}{v_{TA}^3} + \frac{1}{v_{LA}^3} \right) \right]^{-1/3}$$ (S10)

## II. Additional data for the experimental diamond anvil cells

**Table S1.** Additional parameters of the DACs used in the experiments.

| DAC | Culet size (μm) | Gasket | Composition | Pressures (GPa) | Measurement |
|---|---|---|---|---|---|
| #H1 | 100 | W+cBN/epoxy | Ce+NH$_3$BH$_3$ | 88-159 | SC |
| #H2 | 100 | W+cBN/epoxy | Ce+NH$_3$BH$_3$ | 97-140 | SC, XRD |
| #H3 | 100 | W+cBN/epoxy | Ce+NH$_3$BH$_3$ | 95 | SC, XRD |
| #H4 | 80 | W+MgO/epoxy | Ce+NH$_3$BH$_3$ | 86-124 | SC |
| #H5 | 60 | W+MgO/epoxy | Ce+NH$_3$BH$_3$ | 98-166 | SC, XRD |
| #H6 | 50 | W+MgO/epoxy | Ce+NH$_3$BH$_3$ | 174,164,162 | SC, XRD |
| #D1 | 100 | W+MgO/epoxy | Ce+ND$_3$BD$_3$ | 76 | SC, XRD |
| #D2 | 100 | W+MgO/epoxy | Ce+ND$_3$BD$_3$ | 100,120 | SC |
| #D3 | 100 | W+MgO/epoxy | Ce+ND$_3$BD$_3$ | 100 | SC |
| #DH | 100 | W+MgO/epoxy | Ce+ND$_3$BH$_3$ | 88-136 | SC, XRD |
| #S1 | 100 | W | Ce+D$_2$ | 15-140 | XRD |
| #S2 | 100 | W | Ce+ND$_3$BD$_3$ | 29-117 | XRD |
| #S3 | 80 | W | Ce+ND$_3$BD$_3$ | 65,122,128 | XRD |

**Table S2.** Indexed lattice parameters and unit cell volumes of the phases formed in the electrical cells

| Cell | Pressure (GPa) | Phase | Lattice parameter (Å) | | Unit cell volume (Å$^3$) |
|---|---|---|---|---|---|
| #H2 | 137 | $Fm\bar{3}m$-CeH$_{10}$ | a=5.05(0) | | 32.19 |
| | | $P6_3/mmc$-CeH$_9$ | a=3.65(3) | c=5.43(1) | 31.38 |
| | | $I4/mmm$-CeH$_4$ | a=2.77(8) | c=5.72(1) | 22.07 |
| #H3 | 100 | $P6_3/mmc$-CeH$_9$ | a=3.71(3) | c=5.52(5) | 32.98 |
| | | $I4/mmm$-CeH$_4$ | a=2.84(7) | c=5.94(1) | 24.07 |
| #H5 | 150 | $Fm\bar{3}m$-CeH$_{10}$ | a=5.02(0) | | 31.62 |
| | | $P6_3/mmc$-CeH$_9$ | a=3.59(7) | c=5.43(1) | 30.43 |
| | | $I4/mmm$-CeH$_4$ | a=2.74(8) | c=5.73(1) | 21.63 |
| #H6 | 172 | $Fm\bar{3}m$-CeH$_{10}$ | a=4.82(0) | | 30.46 |
| | | $P6_3/mmc$-CeH$_9$ | a=3.55(7) | c=5.37(1) | 29.43 |
| | | $I4/mmm$-CeH$_4$ | a=2.68(0) | c=5.71(0) | 20.50 |

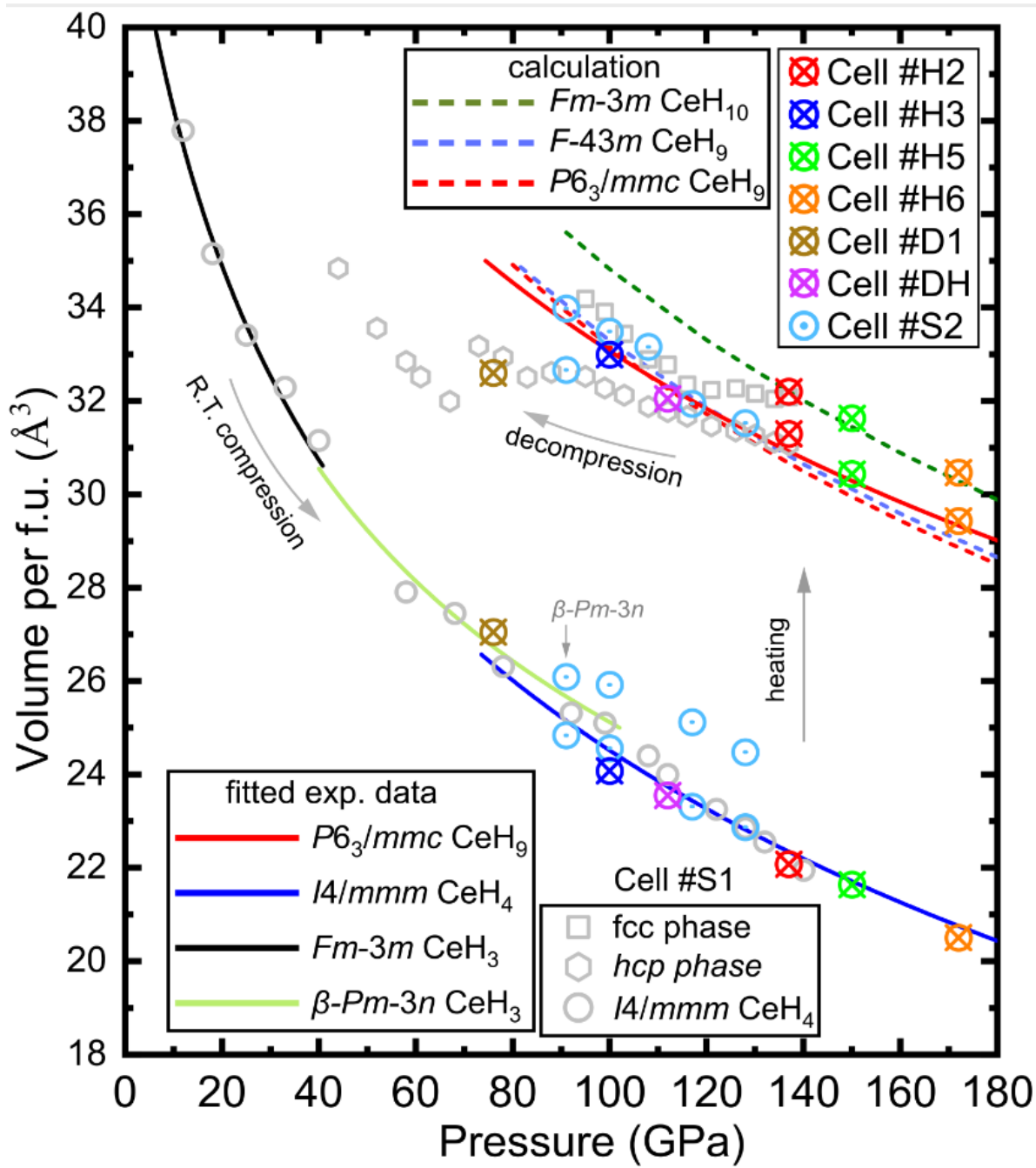

**Fig. S1.** Pressure dependence of the unit cell volume (per f.u.) of cerium hydrides. Experimental data in this study are represented by different symbols. Dashed and solid lines indicate the calculated *P–V* data and fitted experimental data from Ref.[24,25], respectively.

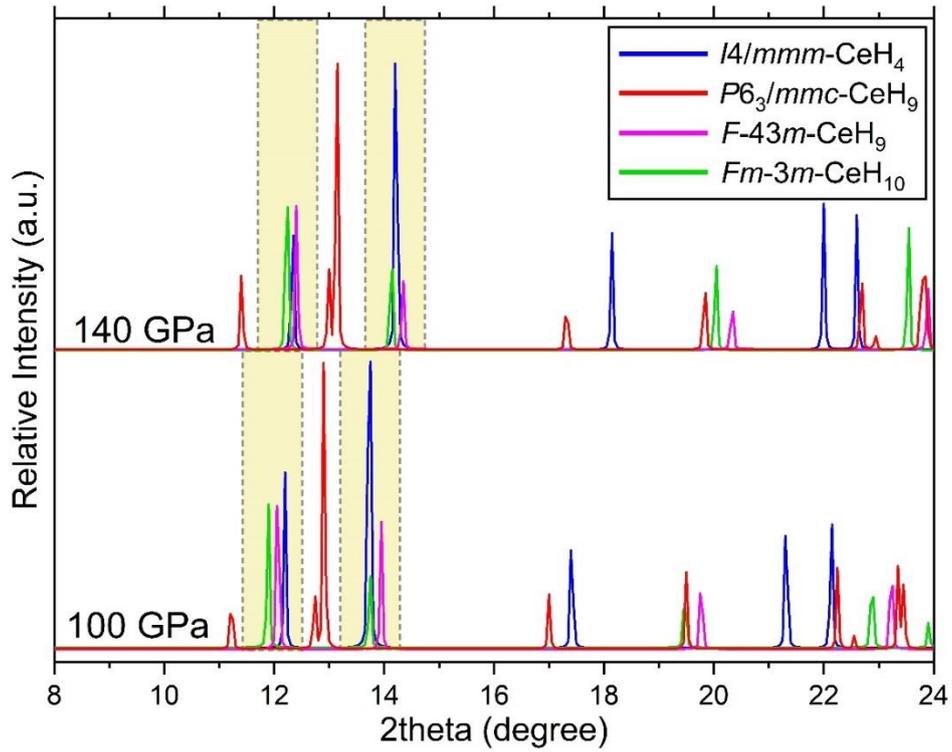

**Fig. S2.** Calculated diffraction patterns for different phases at 100 and 140 GPa ($\lambda$ = 0.6199 Å).

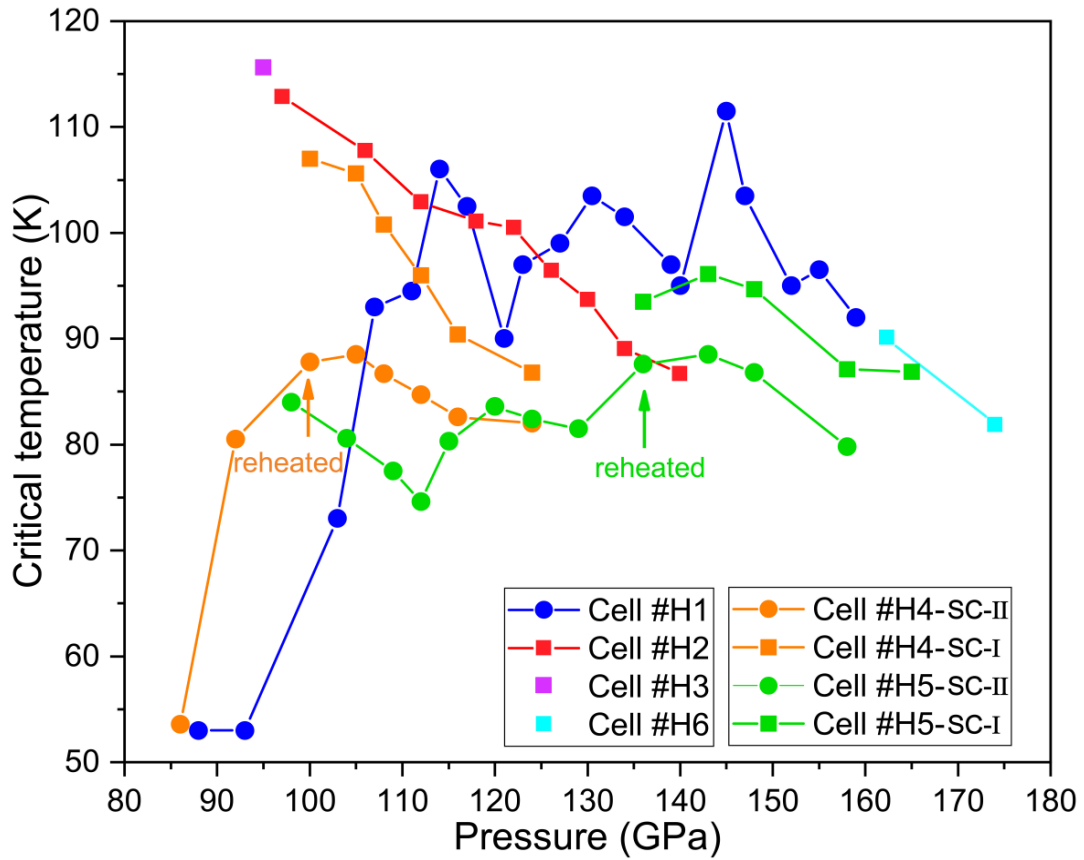

**Fig. S3** $T_C$–$P$ dependence for $Fm\bar{3}m$-CeH$_{10}$ (square) and $P6_3/mmc$-CeH$_9$ (circle) together. Data for the cell #H1 are the average $T_C$ of cooling and warming cycles.

**Cell #H1**

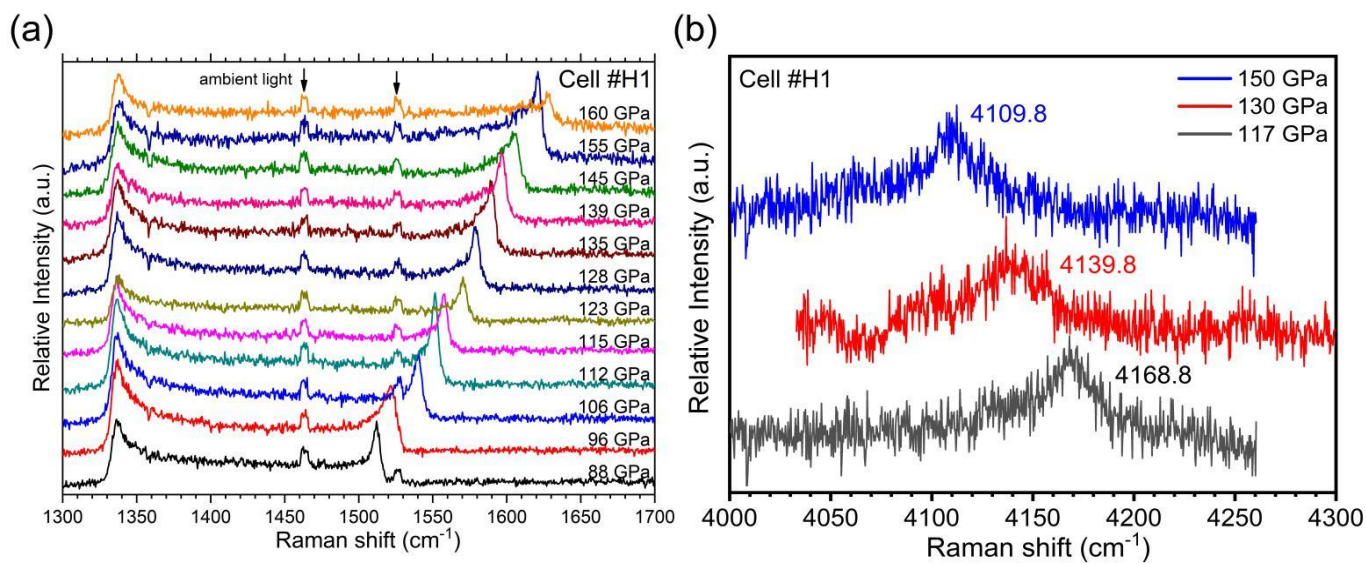

**Fig. S4.** Raman spectra of the sample in cell #H1. (a) Pressure measurements using the Raman shift of diamond. (b) Raman signal from the $H_2$ vibration.

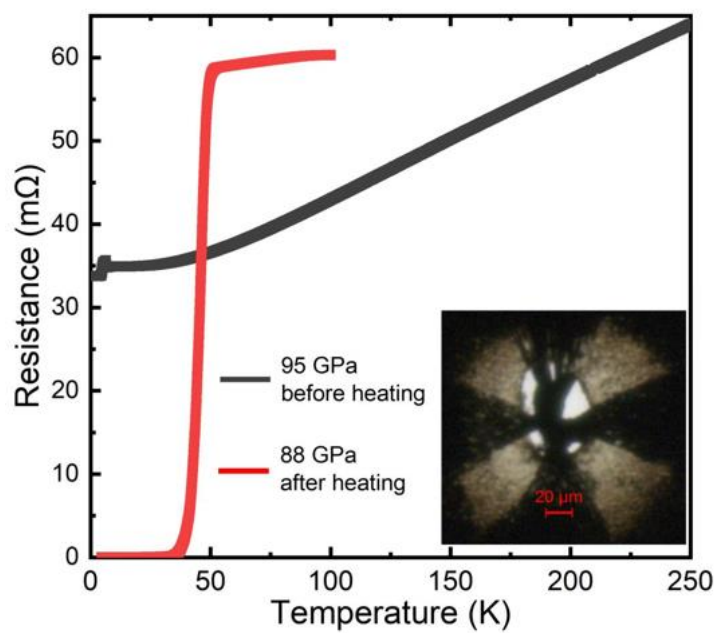

**Fig. S5.** Temperature dependence of the resistance in the cooling cycle in cell #H1 before and after the laser heating. Inset: the DAC chamber photographed in transmitted light.

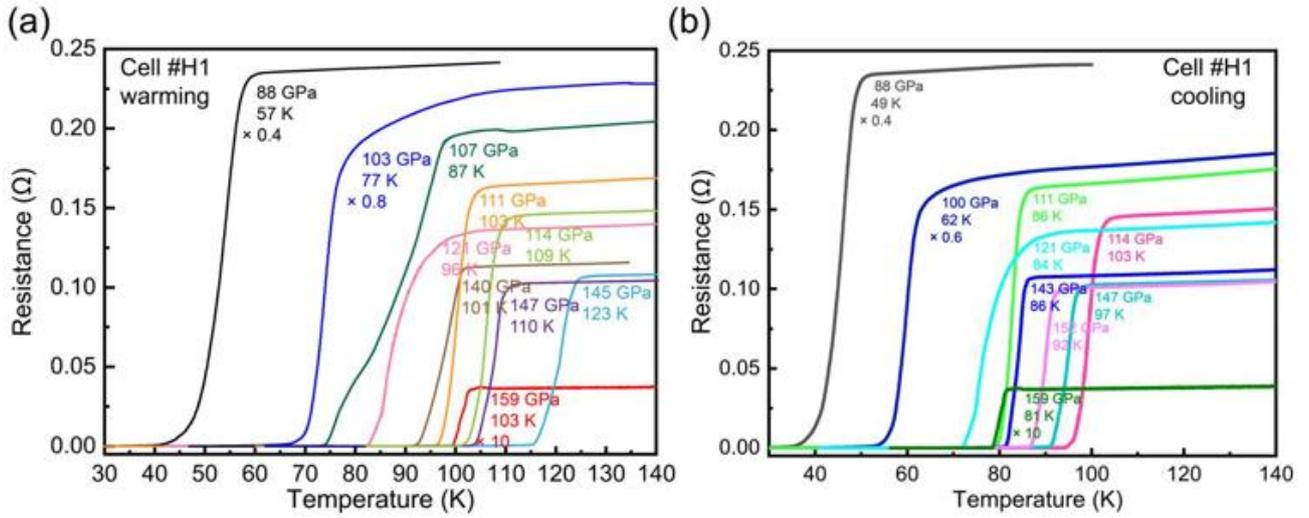

**Fig. S6.** Superconducting transitions in the sample in cell #H1 during (a) warming and (b) cooling cycles.

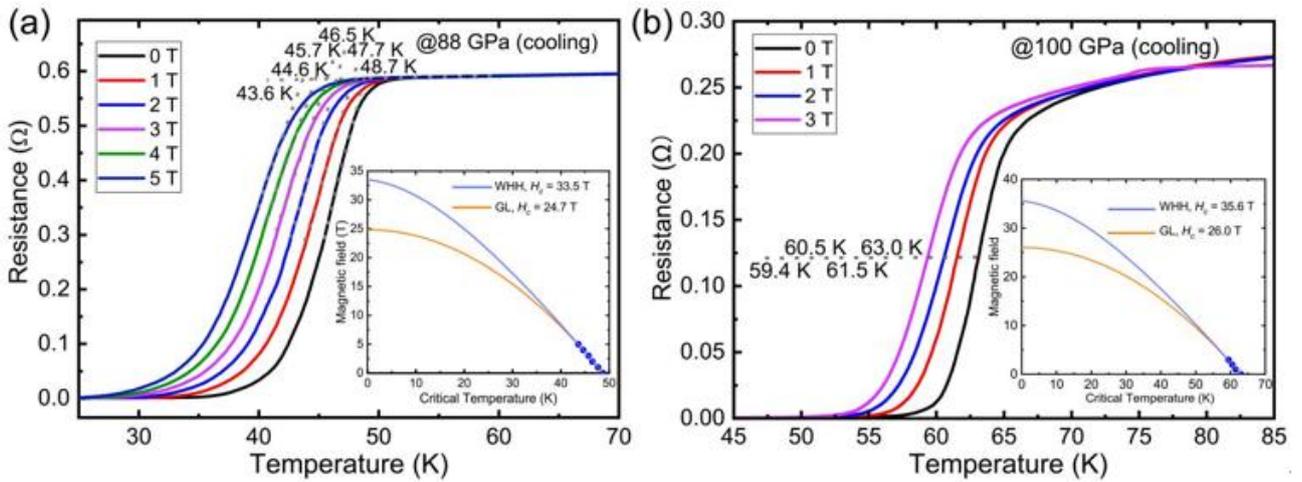

**Fig. S7.** Superconducting transitions of CeH$_9$ in an external magnetic field during the cooling cycle. (a) In a magnetic field of 0–5 T at 88 GPa. Inset: Upper critical magnetic field estimated using the WHH and GL theories. $T_C$ is defined at the onset of the resistance drop. (b) In a magnetic field of 0–3 T at 100 GPa. Inset: Upper critical magnetic field estimated using the WHH and GL theories. $T_C$ is defined when the resistance drops to half of the normal level because the beginning of the transition is hard to recognize.

**Cell #H2**

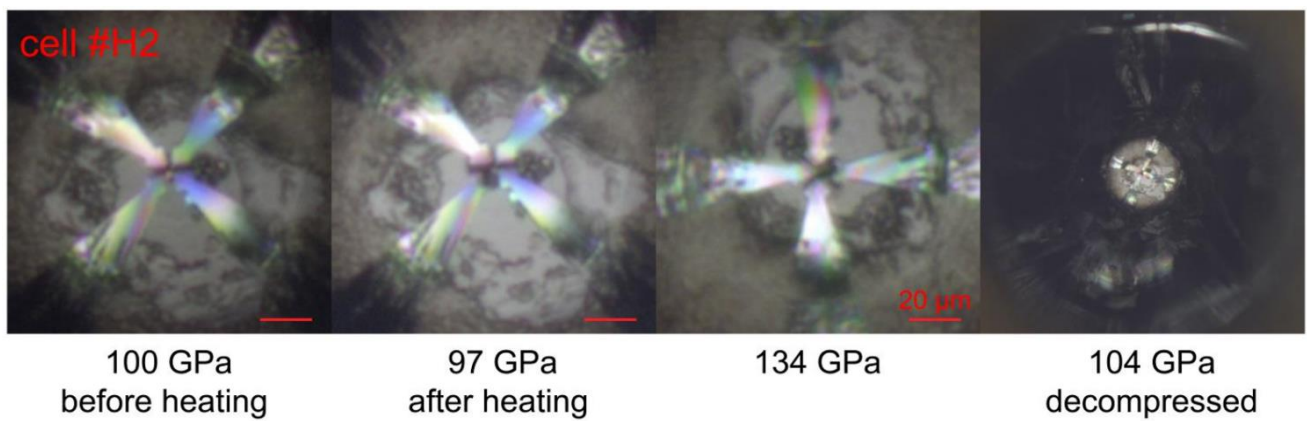

| 100 GPa | 97 GPa | 134 GPa | 104 GPa |
| before heating | after heating | | decompressed |

**Fig. S8.** Cell #H2 photographed at selected pressures.

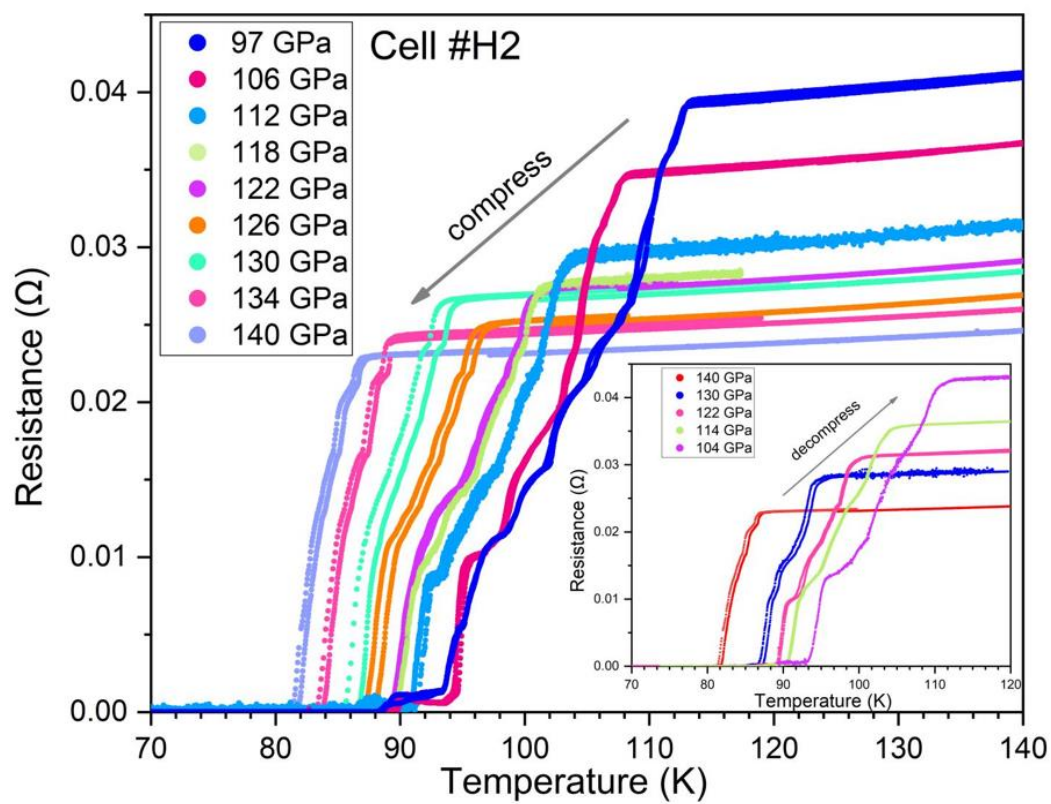

**Fig. S9.** Superconducting transitions in the CeH$_x$ sample in cell #H2 during compression and decompression (inset). The warming and cooling R(T) curves almost coincide at the same pressure.

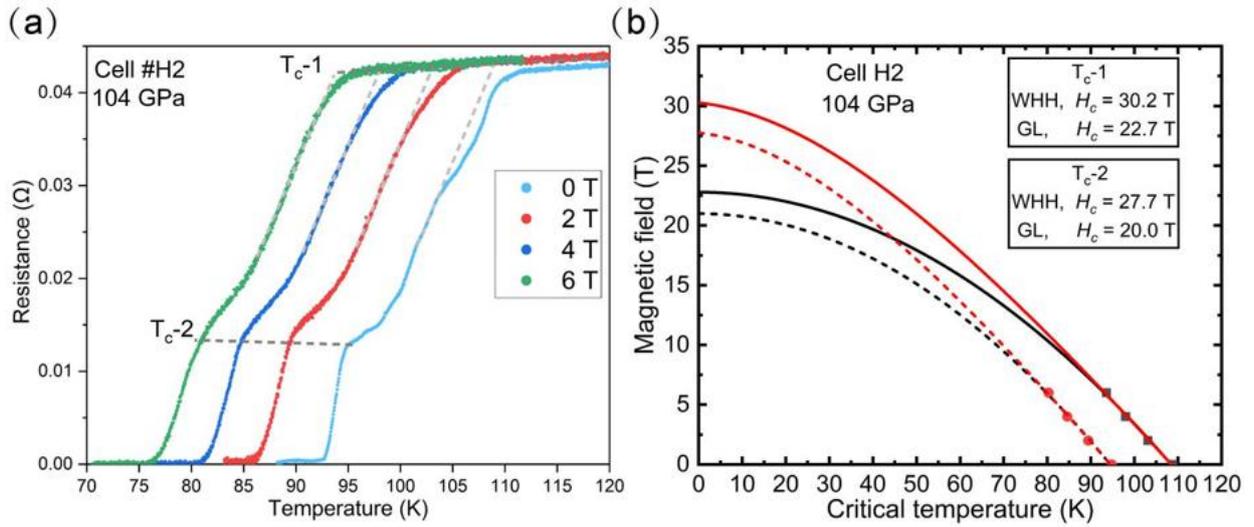

**Fig. S10.** Superconducting transitions in the CeH$_x$ sample in cell #H2 in an external magnetic field at 104 GPa. (a) $R$–$T$ curves with two-stage transitions. (b) Upper critical field estimated using the WHH and GL theories.

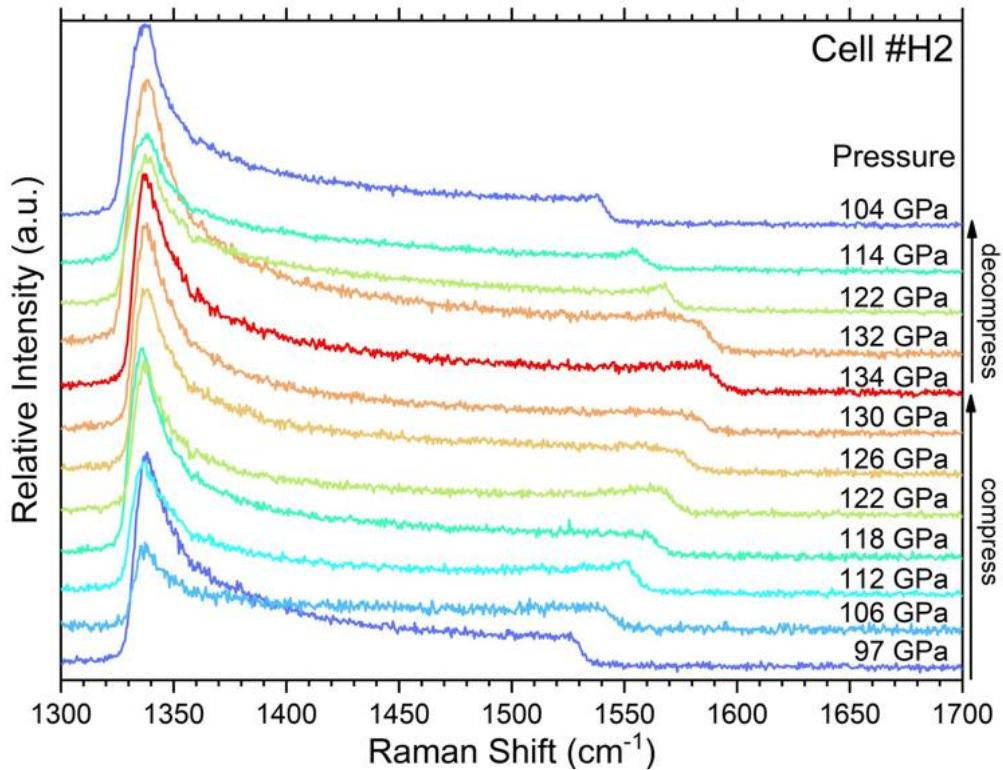

**Fig. S11.** Pressure in cell #H2 measured using the Raman shift of diamond.

## Cell #H3

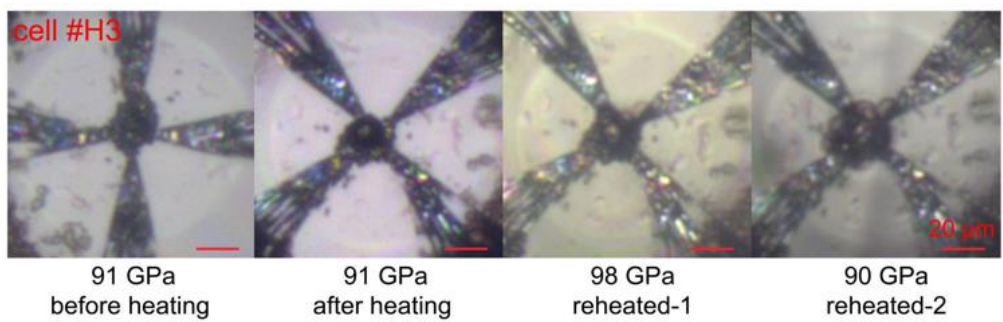

91 GPa
before heating

91 GPa
after heating

98 GPa
reheated-1

90 GPa
reheated-2

**Fig. S12.** Cell #H3 photographed at selected pressures.

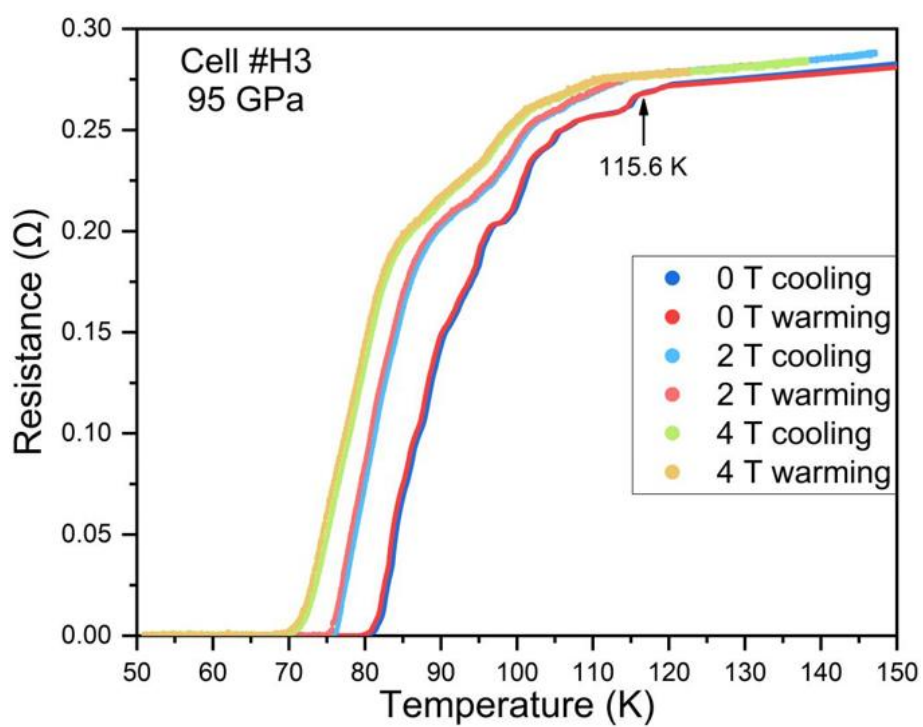

**Fig. S13.** Superconducting transitions in the sample in cell #H3 in an external magnetic field at 95 GPa.

**Cell #H4**

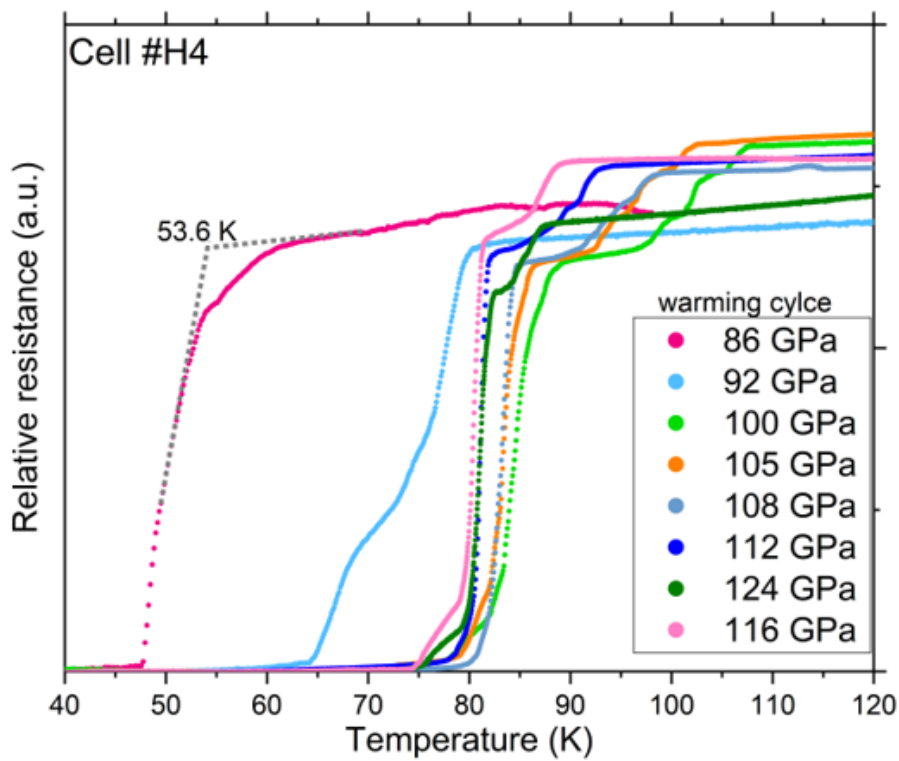

**Fig. S14.** Superconducting transitions in the sample in cell #H4.

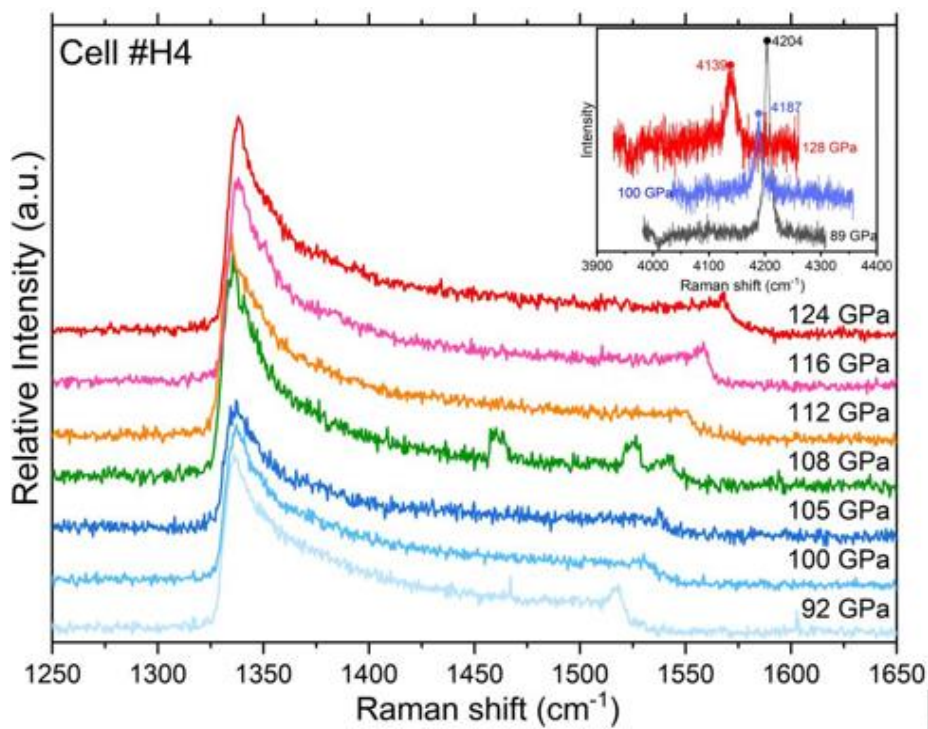

**Fig. S15.** Pressure measurements in cell #H4 using the first-order Raman signal of diamond. The Raman signal of $H_2$ is shown in the inset.

**Cell #H5**

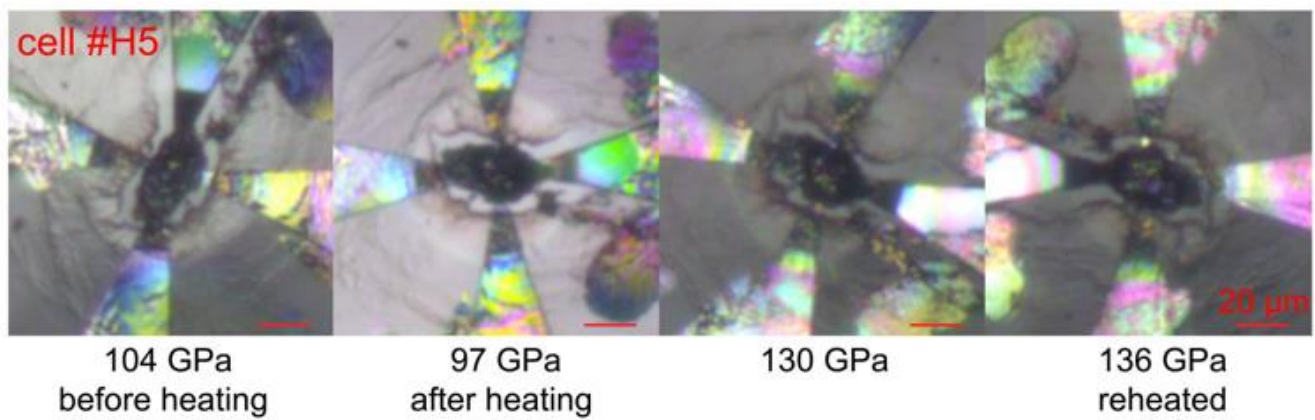

| 104 GPa | 97 GPa | 130 GPa | 136 GPa |
| before heating | after heating | | reheated |

**Fig. S16.** Cell #H5 at selected pressures.

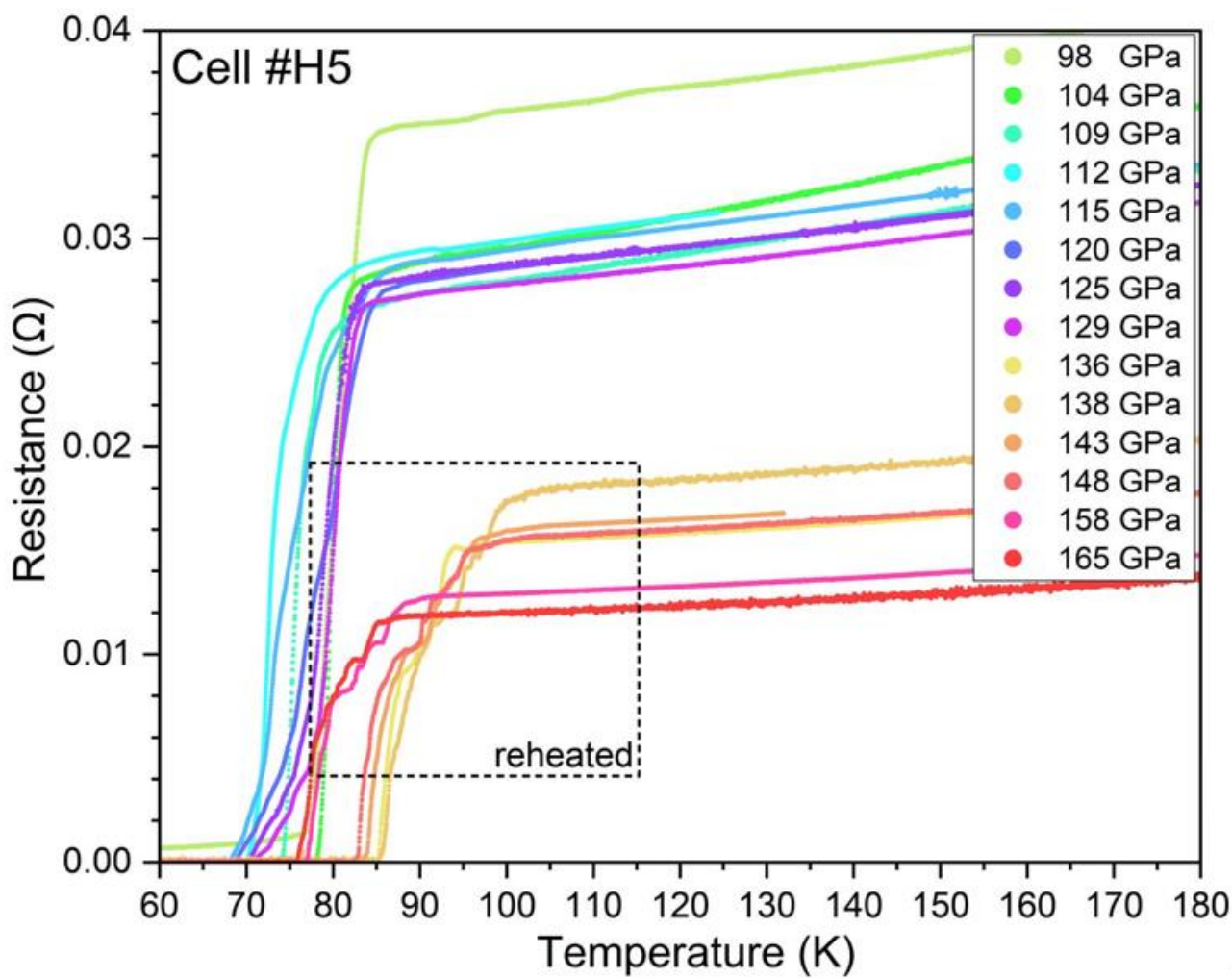

**Fig. S17.** Superconducting transitions in the sample of cell #H5. Dashed box marks the transitions after reheating of the sample.

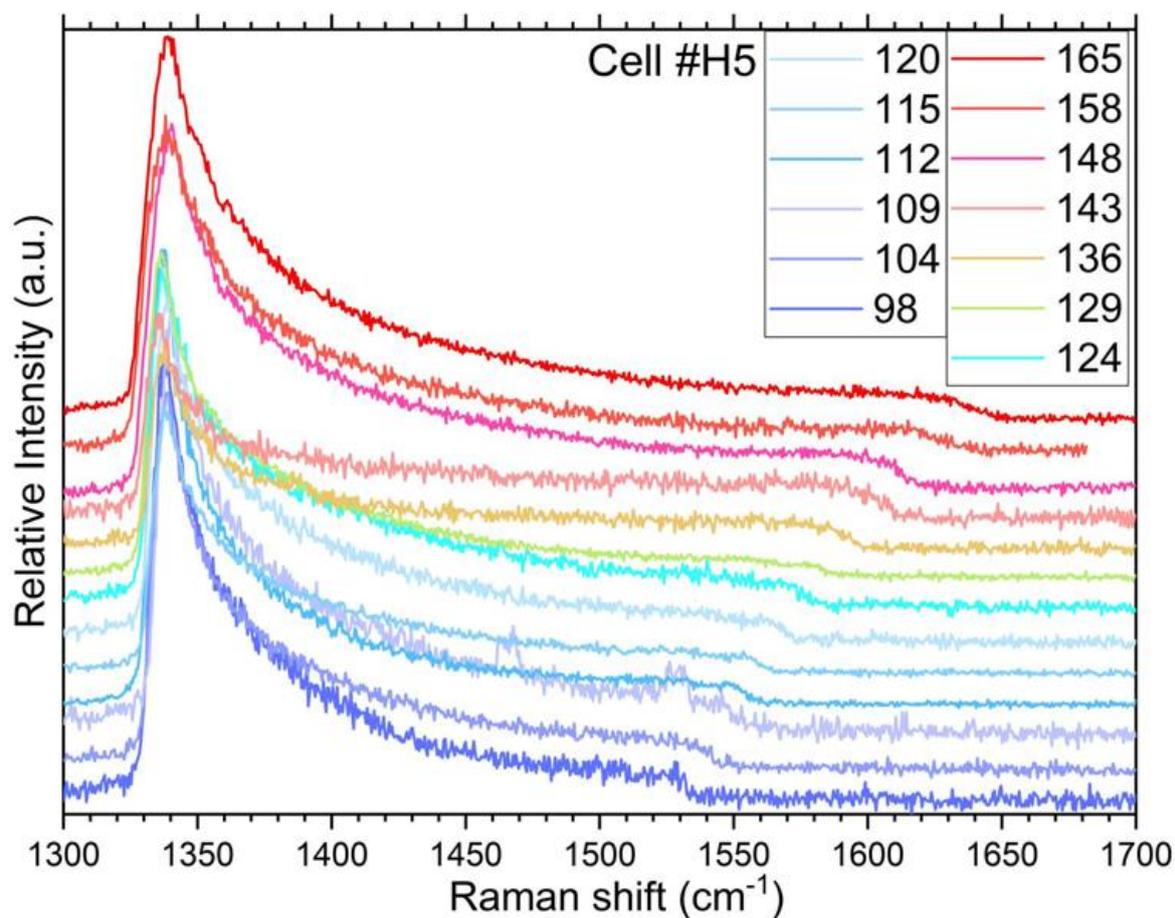

**Fig. S18.** Pressure measurements in the sample in cell #H5 using the Raman shift of diamond.

**Cell #H6**

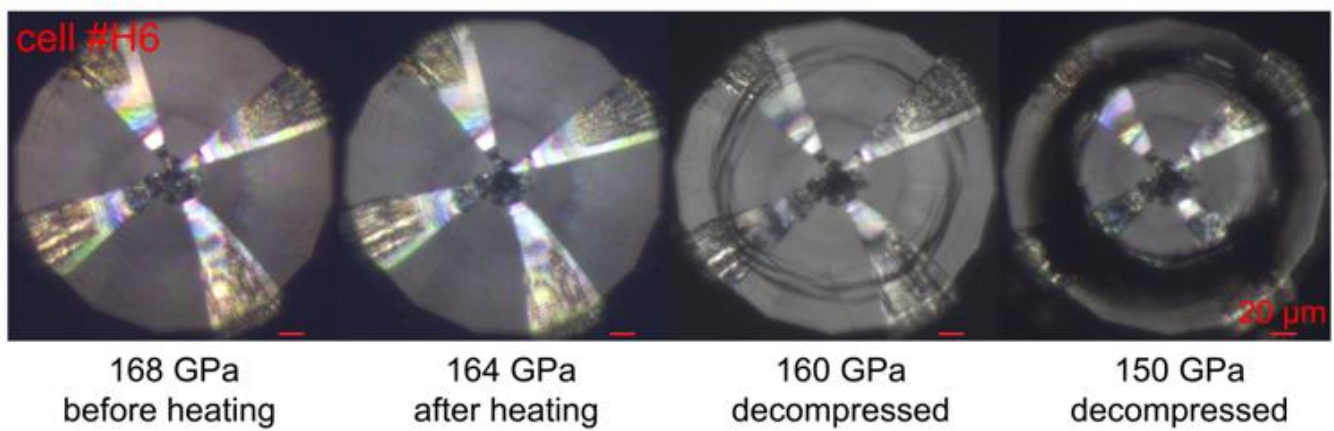

168 GPa before heating          164 GPa after heating          160 GPa decompressed          150 GPa decompressed

**Fig. S19.** Cell #H6 photographed at selected pressures.

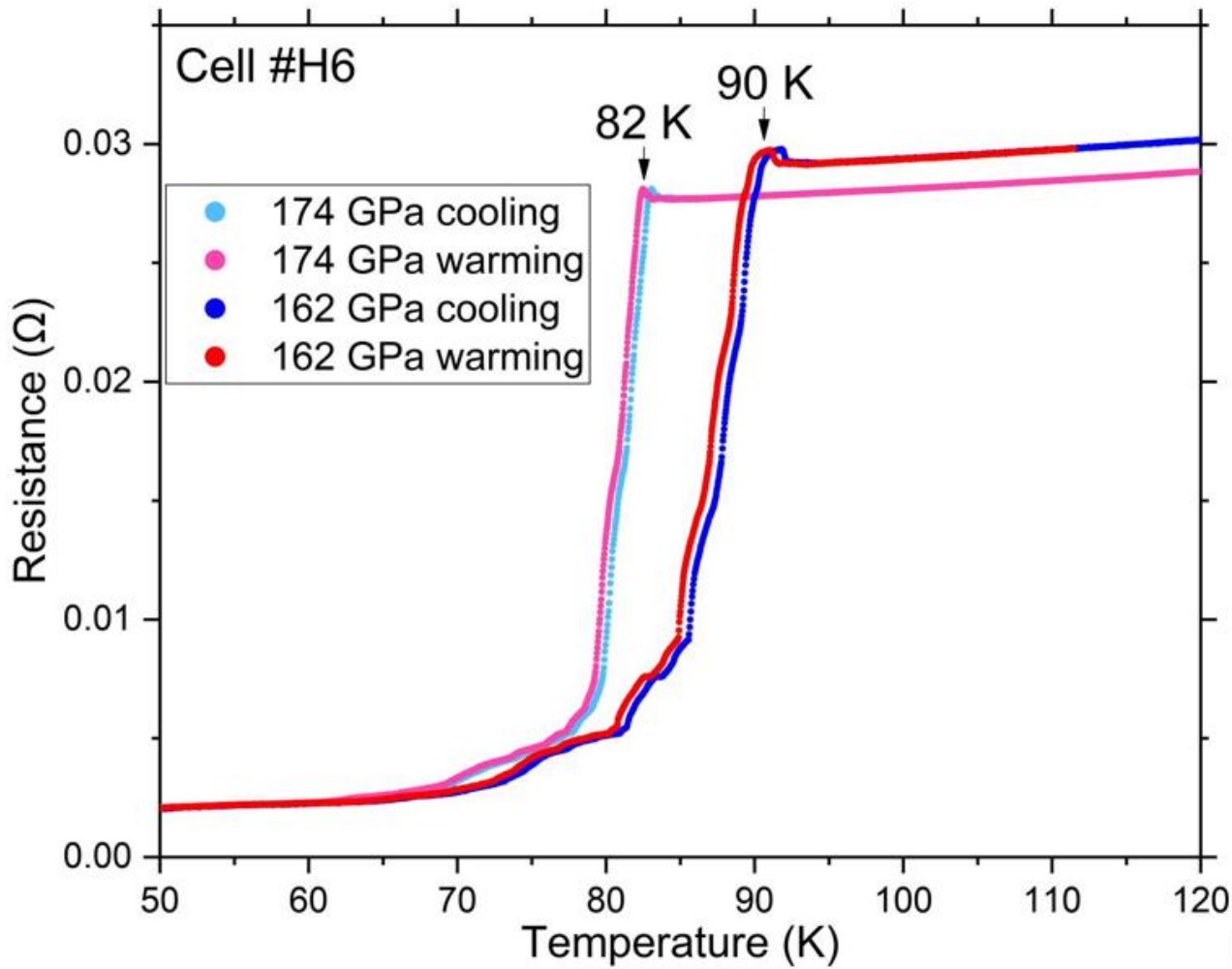

**Fig. S20.** Superconducting transitions in the CeH$_x$ sample in cell #H6 at 162 and 174 GPa.

**Cell #S1**

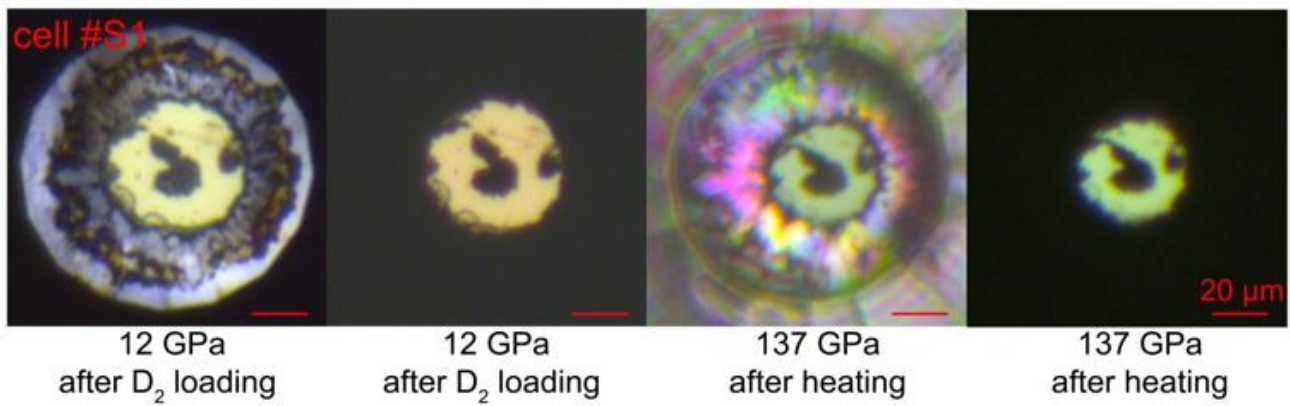

**Fig. S21.** Cell #S1 photographed at selected pressures.

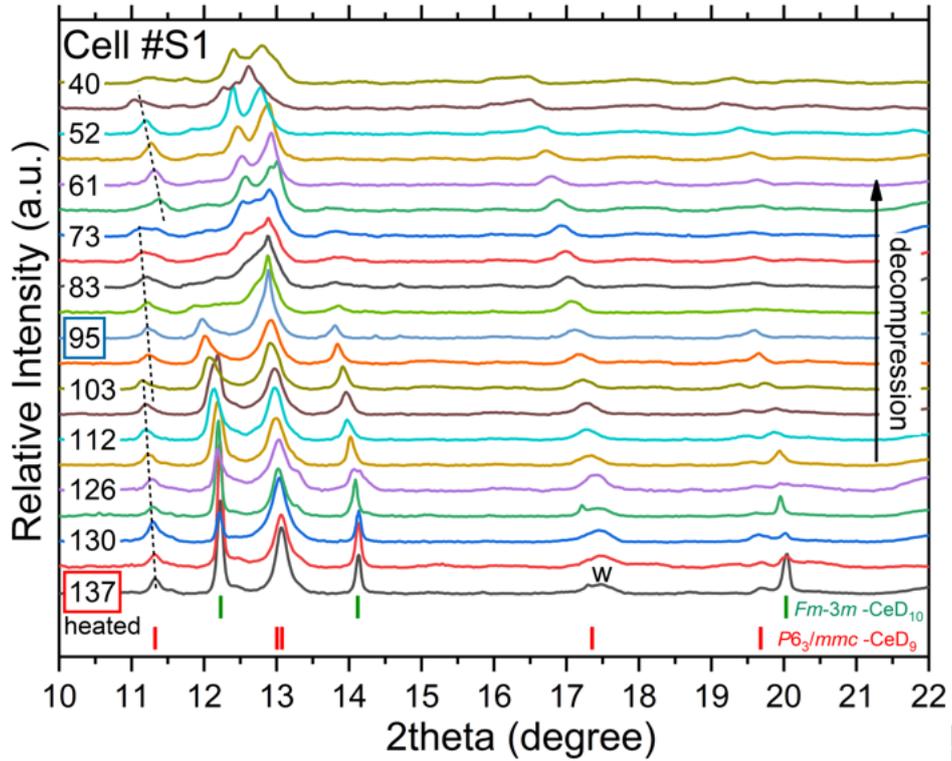

**Fig. S22.** X-ray diffraction patterns ($\lambda = 0.6199$ Å) of the $CeD_x$ sample in cell #S1 (heated at 137 GPa) during decompression. The pressure values are indicated on the left side in GPa.

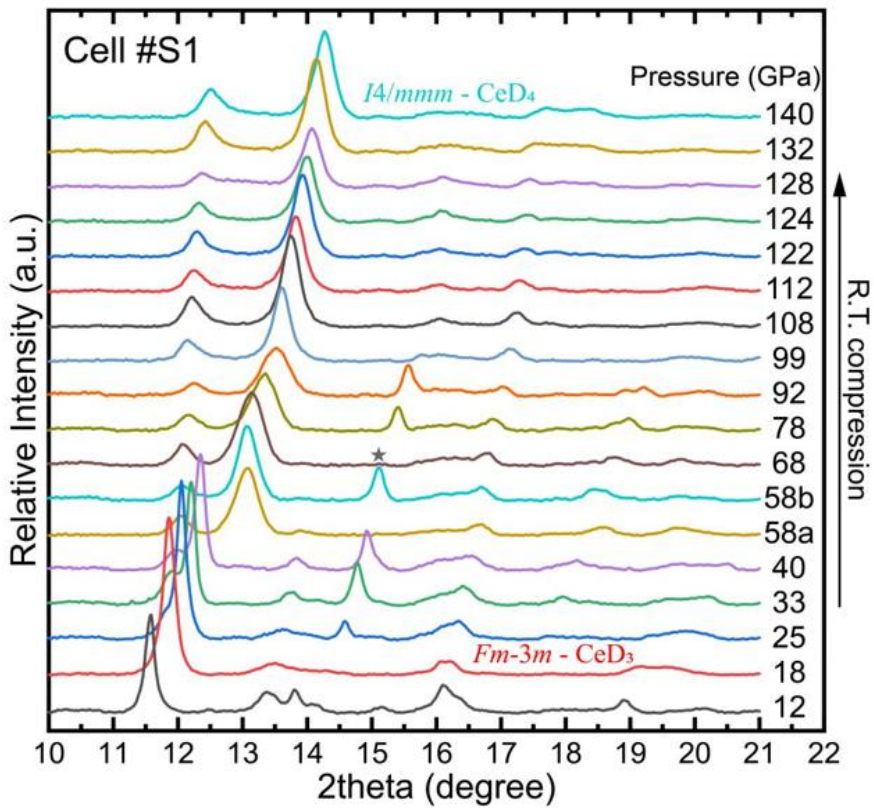

**Fig. S23.** X-ray diffraction patterns ($\lambda = 0.6199$ Å) of the $CeD_x$ sample in cell #S1 during compression at room temperature. The pressure values are indicated on the right.

**Cell #S2**

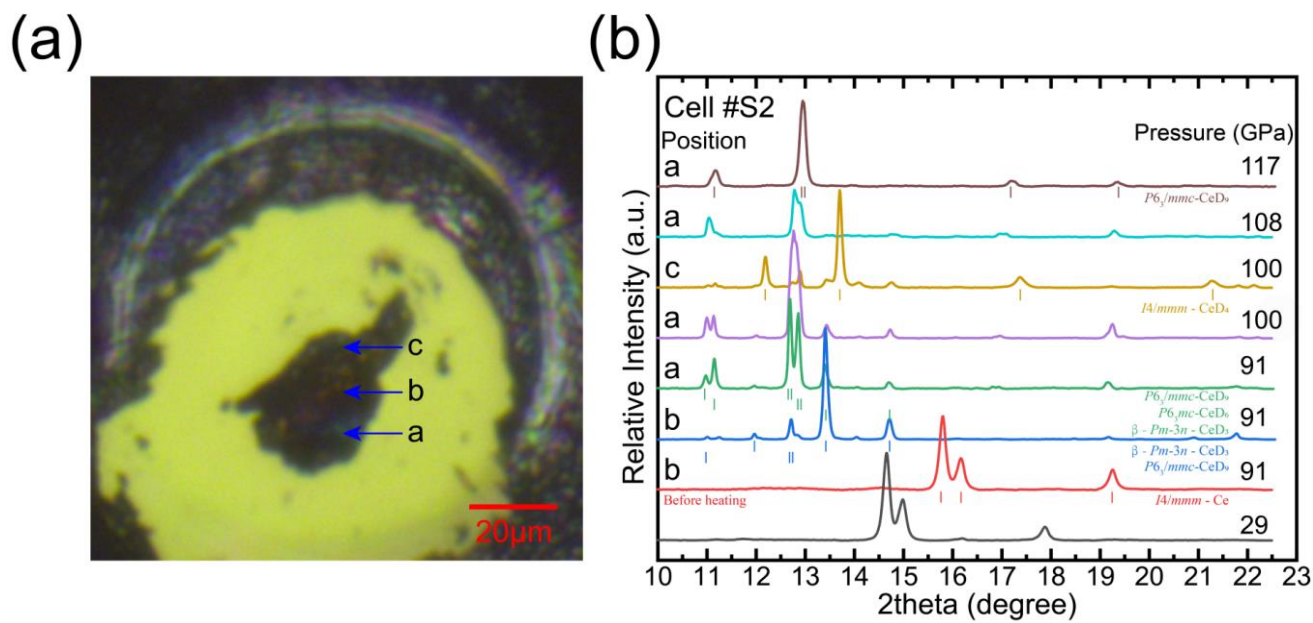

**Fig. S24.** X-ray diffraction of the CeD$_x$ sample in cell #S2. (a) Photograph of cell #S2 at 91 GPa after laser heating with marks (a,b,c) of the investigated regions of the sample. (b) X-ray diffraction patterns ($\lambda = 0.6199$ Å) of the CeD$_x$ sample at selected pressures and different positions.

**Cell #S3**

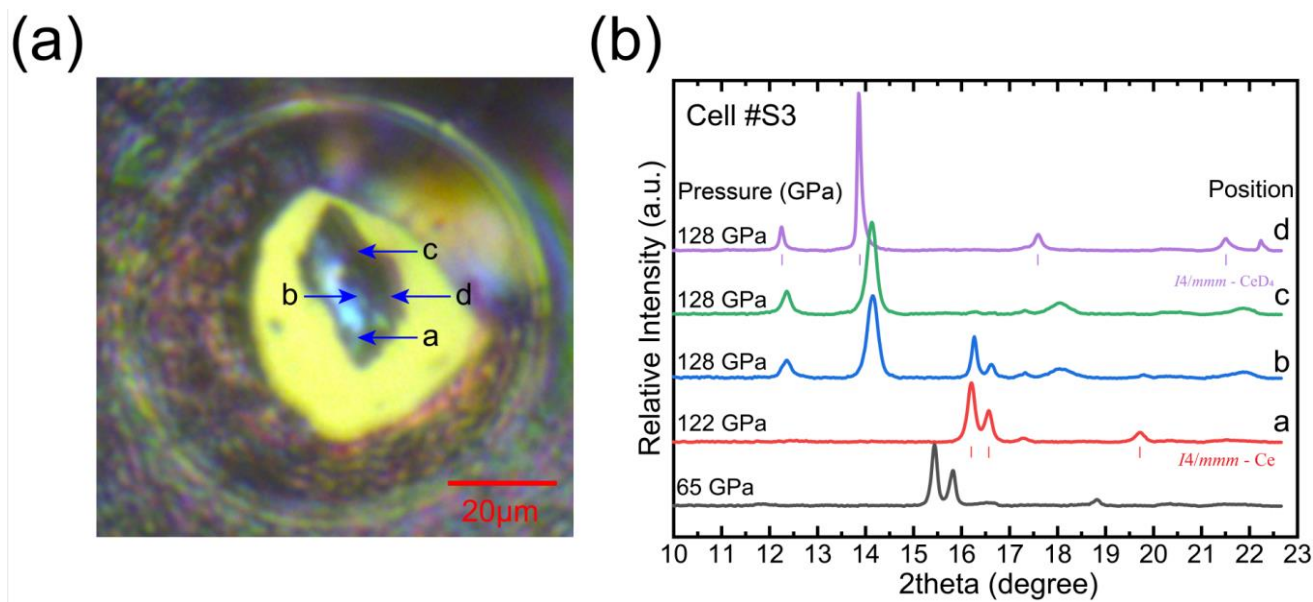

**Fig. S25.** X-ray diffraction of the CeD$_x$ sample in cell #S3. (a) Photograph of cell #S3 at 128 GPa after laser heating with marks (a,b,c,d) of the investigated positions. (b) X-ray diffraction patterns of the CeD$_x$ sample at selected pressures and different positions ($\lambda = 0.6199$ Å).

**Cell #D1**

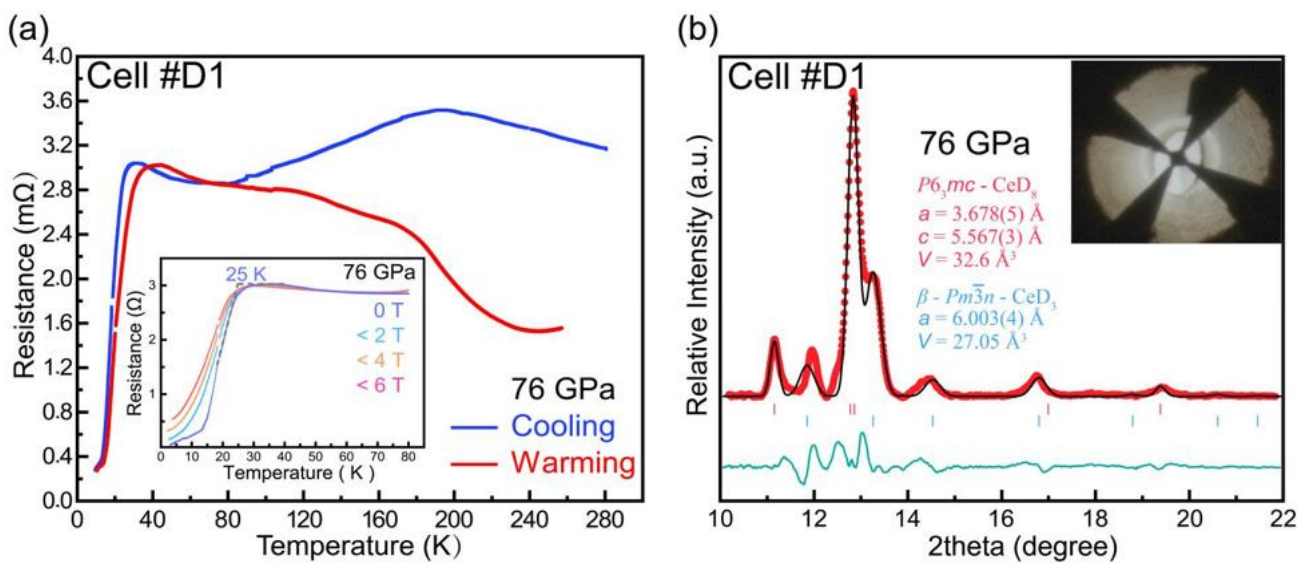

**Fig. S26.** Superconducting properties of the cerium deuterides synthesized in cell #D1. (a)Superconducting transitions of the CeD$_x$ sample in cell #D1 at 76 GPa. The transitions in an external magnetic field are shown in the inset. (b) Le Bail refinement of the unit cell parameters of CeD$_8$ and CeD$_3$ synthesized in cell #D1 at 76 GPa.

**Cell #D2**

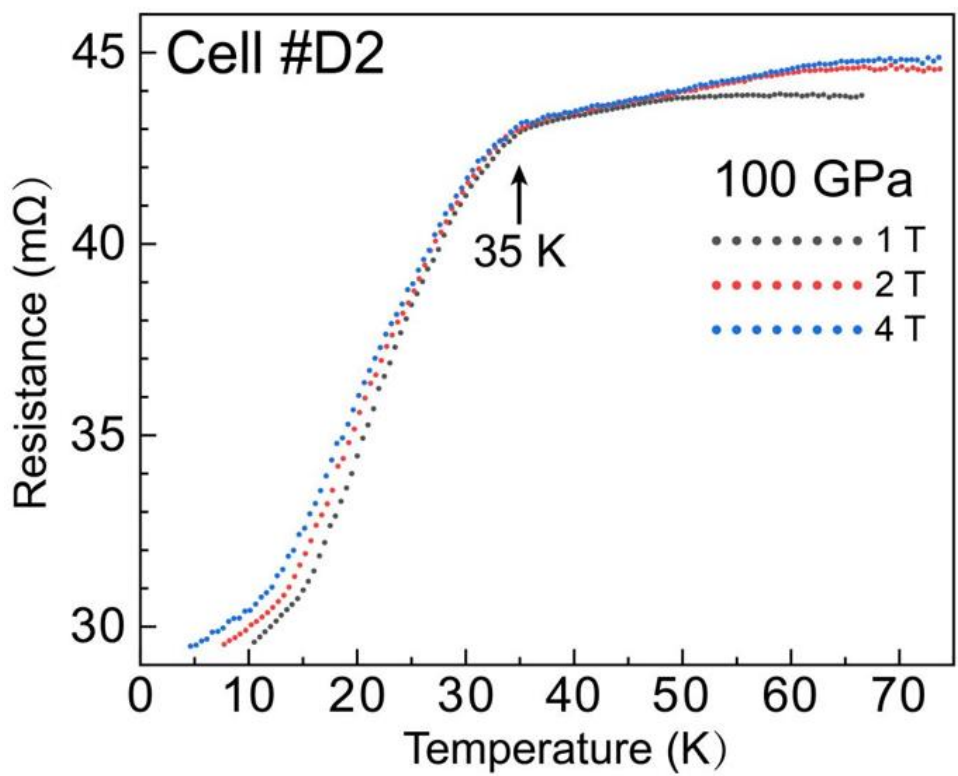

**Fig. S27.** Superconducting transitions in the CeD$_x$ sample in cell #D2 in an external magnetic field at 100 GPa.

**Cell #D3**

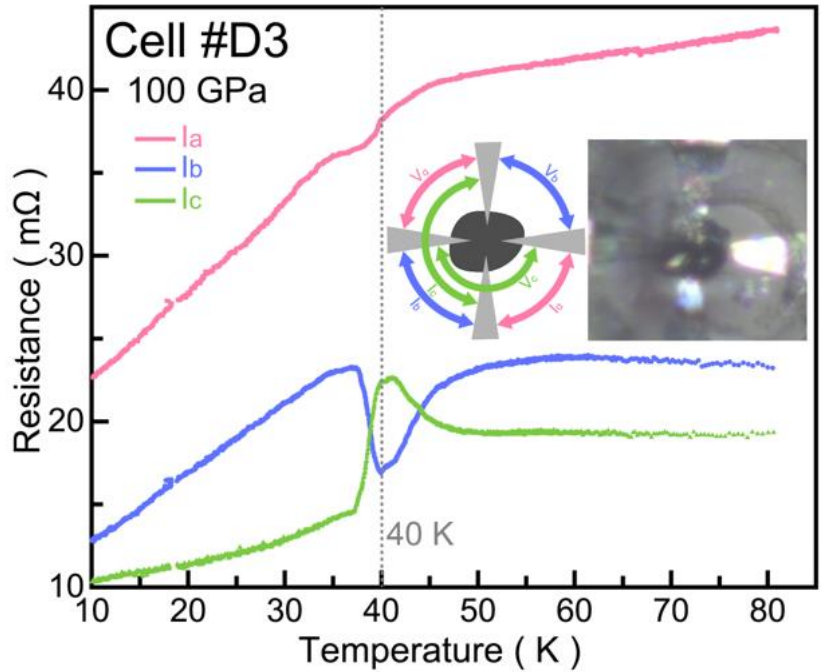

**Fig. S28.** Temperature dependence of the electrical resistance of the CeD$_x$ sample in cell #D3 measured using different combinations of electrodes at 100 GPa. Insets are the schematic diagram of the 4-probe electric measurements and the photograph of the sample.

**Cell #DH**

In order to study the partial replacement of hydrogen by deuterium in the structure of CeX$_9$ we loaded a cerium particle with ND$_3$BH$_3$ (which generates HD when heated) to the DAC #DH and compressed it to about 90 GPa. After laser heating the pressure in this cell decreased to 88 GPa. We conducted the electrical measurements and noticed that the resistance began to drop to zero at 60 K (88 GPa). This is consistent with the superconducting transition of CeH$_9$, thus we believe that hydrogen reacts with Ce easier than deuterium, and CeH$_9$ is formed first of all.

After we reheated the sample above 1500 K, the pressure increased which resulted in an increase in $T_C$. The XRD pattern at 112 GPa shows that the sample is a mixture of $P6_3/mmc$-Ce(H,D)$_9$ and $I4/mmm$-Ce(H,D)$_4$ (Fig. S29). Further compression destroyed one of the electrodes, and we continued with two remaining ones. Measured $T_C$(P) of the sample is lower compared with that of CeH$_9$ at the same conditions (Fig. S30a). The superconducting nature of the observed transitions was confirmed with an applied external magnetic field (Fig. S30b). As can be seen in the Figure S30a, some superconducting transitions have one or two steps (e.g., at 105 GPa) which may correspond to the presence of partially deuterated compounds: $P6_3/mmc$ -CeH$_7$D$_2$, CeH$_5$D$_5$ etc. These steps, however, are not sufficiently pronounced to be used in calculations of the isotope effect or the degree of substitution of hydrogen by deuterium.

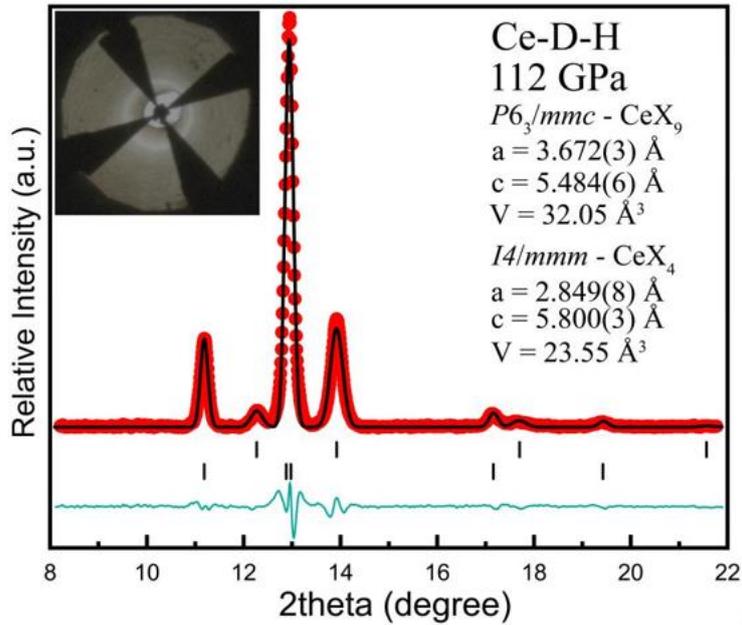

**Fig. S29.** Le Bail refinement of the experimental XRD pattern of the sample at 112 GPa in cell #DH. X represents H or D. The chamber photographed in transmitted light is shown in the inset.

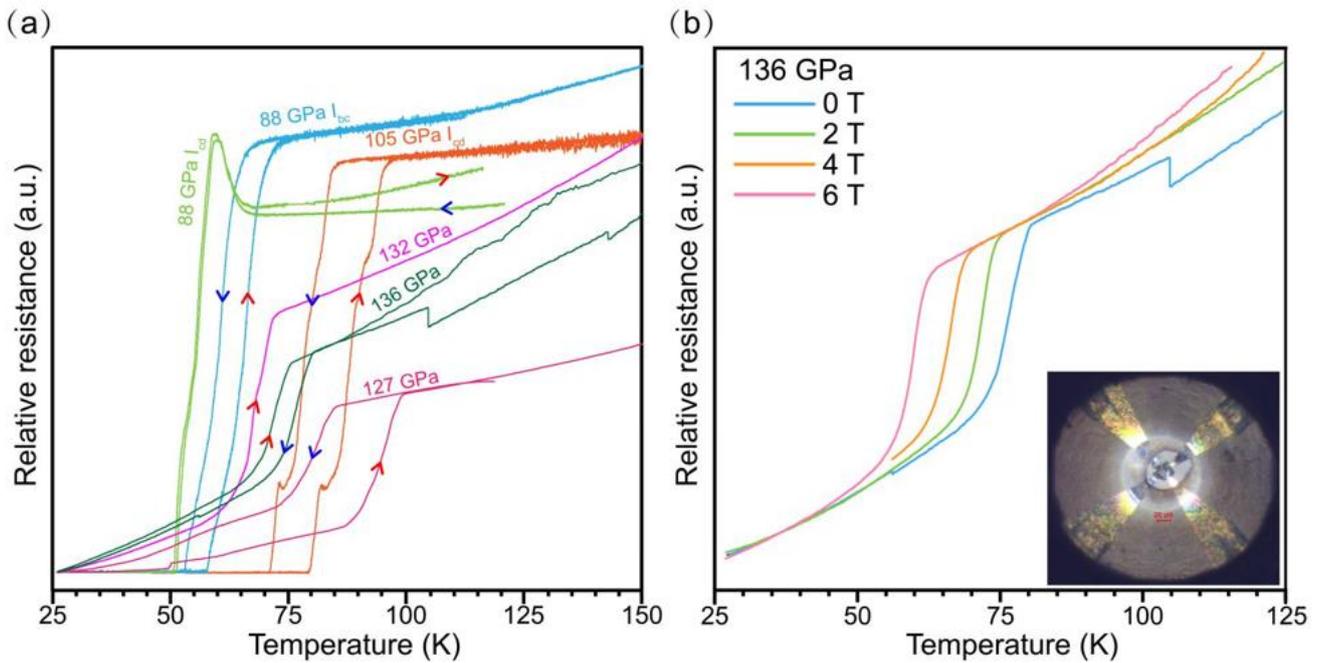

**Fig. S30.** Superconducting transitions in the Ce–H–D sample in cell #DH. Some curves are moved vertically for an easier comparison. (a) Temperature dependence of the resistance at selected pressures. Blue and red arrows represent cooling and heating cycles. (b) Temperature dependence of the resistance in an external magnetic field at 136 GPa. Inset: a photograph of the DACs #DH chamber.

# III. Ab initio calculations of superconducting parameters of CeH₉ at 120-200 GPa

Here we discuss the discrepancies between the experimental and predicted $T_C$ ($P6_3/mmc$-CeH$_9$) at 120 – 200 GPa. Using both the interpolation[13] and the optimized tetrahedron[12] methods at pressures of 100, 120, 150, and 200 GPa, we found that the pressure dependence of the critical temperature $T_C(P)$ should have a well-known dome-like shape (Fig. 4d): above 120 GPa the EPC coefficient $\lambda$ of $P6_3/mmc$-CeH$_9$ (SC-II) decreases monotonically from 1.46, reaching 0.68 at 200 GPa. An increase in the logarithmically averaged frequency $\omega_{log}$ and the Debye temperature with a simultaneous decrease in the critical temperature and the EPC coefficient was confirmed by both calculations in Quantum ESPRESSO and calculations of elastic tensors in VASP (Table S9). The $\omega_{log}$ drops by a half from 1272 K to 650 K as the pressure decreases from 200 to 120 GPa because of a low-frequency shift of a large group of phonon modes in $P6_3/mmc$-CeH$_9$ (Fig. S30) which leads to distortion of the hexagonal lattice and a sharp increase in $\lambda$ and $T_C$, followed by the phase transition $P6_3/mmc \rightarrow C2/c$.

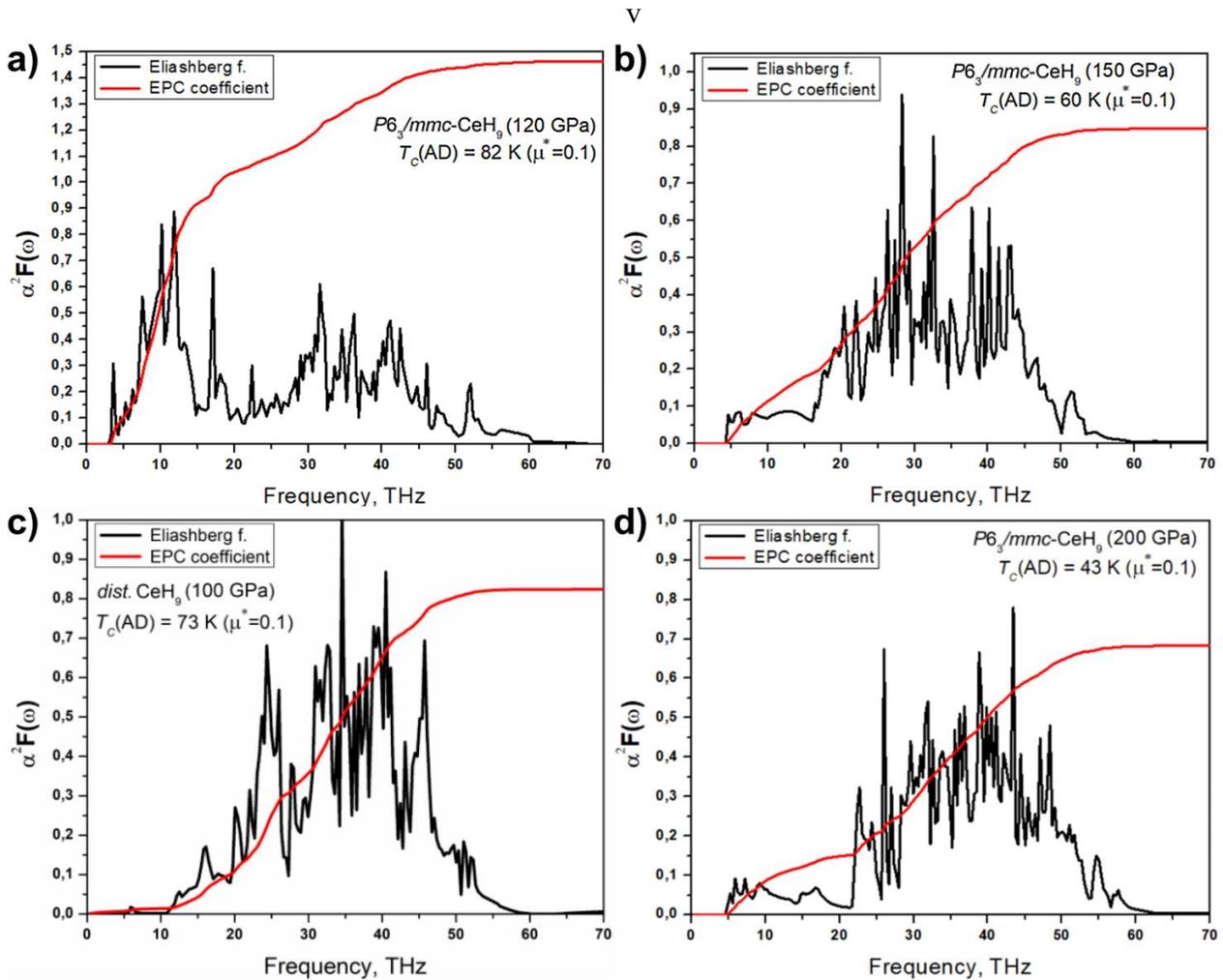

**Fig. S31.** Isotropic Eliashberg functions for $P6_3/mmc$-CeH$_9$ calculated within the optimized tetrahedron method at (a) 120 GPa, (b) 150 GPa, (c) 100 GPa (distorted structure), and (d) 200 GPa.

The experimental data for pressures about 150-200 GPa, however, point to some discrepancies with the predicted value of $T_C$. Within both the interpolation and optimized tetrahedron methods using the PBE[26] and GHHT[10,11] pseudopotentials that give consistent results, calculated $T_C$ in all cases was found to be lower (43–60 K at the Coulomb pseudopotential $\mu^* = 0.1$) than the experimental value (>80 K). This fact may be explained by the presence of impurities of cubic $F\bar{4}3m$-CeH$_9$ and $Fm\bar{3}m$-CeH$_{10}$ phases[27]. Another likely solution lies in the significant anisotropy of the superconducting gap of hexagonal CeH$_9$ as it was observed for CaC$_6$[28] and MgB$_2$[29]. The detailed analysis of partial EPC coefficients $\lambda_{qv}$ shows that in some directions the electron–phonon interaction is more intensive and $\lambda$ reaches 1.2–1.44 at 150 GPa, which corresponds to the upper bound of the superconducting gap of ~20 meV, whereas average $\Delta_{iso} \sim 10$ meV. However, this effect may not be too large because in most cases the studied samples are mixtures of randomly oriented nanocrystals. It is known[30] that a random orientation of crystals in MgB$_2$ reduces the critical transition temperature from 40 to 32–34 K, making it closer to the results of isotropic calculations.

Additional factor of an increase in $T_C$(CeH$_9$) may be a reduced value of the Coulomb pseudopotential $\mu^* = 0.05$–0.1 as shown in Table S3 (values in parentheses). Note that a contradiction with the data of Salke et al.[25] at 200 GPa is likely due to the use of the minimum smoothing $\sigma_1 = 0.005$ Ry, which in our calculations also gives $T_C$ ~100 K. However, the convergence is achieved only with a further increase in $\sigma$, and the converged $T_C$ is almost two times lower, in agreement with the optimized tetrahedron method.

**Table S3.** Parameters of the superconducting state of $P6_3/mmc$-CeH$_9$ at different pressures calculated using the isotropic Migdal–Eliashberg equations (E)[31] and the Allen–Dynes formula (A–D)[16] with the Coulomb pseudopotential $\mu^* = 0.15$–0.1 and 0.05 (shown in parentheses).

| Parameter | 100 GPa | 120 GPa | 150 GPa | 200 GPa |
|---|---|---|---|---|
| $\lambda$ | 0.82 | 1.46 | 0.85 | 0.68 |
| $\omega_{log}$, K | 1393 | 650 | 1084 | 1272 |
| $\omega_2$, K | 1588 | 1022 | 1392 | 1611 |
| $\alpha$ | 0.41-0.47 (0.49) | 0.46-0.48 (0.49) | 0.41-0.47 (0.49) | 0.37-0.46 (0.49) |
| $T_C$ (A–D), K | 51-73 (96) | 68-82 (99) | 42-60 (80) | 26-43 (64) |
| $T_C$ (E), K | 62-80 (102) | 76-87 (101) | 49-64 (82) | 31-46 (65) |
| $N(E_F)$, states/eV·Ce | 0.65 | 0.920 | 0.823 | 0.739 |
| $T_C$ (CeD$_9$), K | 45-72 | 61-71 | 46-59 | 23-34 |
| $\Delta(0)$, meV | 10-13.2 (17.5) | 14.4-17.1 (21) | 7.9-10.6 (14) | 4.8-7.3 (10.6) |
| $\mu_0 H_C(0)$, T | 11.2-14.6 (18.9) | 21-24.5 (28.9) | 10.4 -13.7 (18) | 5.8-8.7 (12.5) |
| $\Delta C/T_C$, mJ/mol·K$^2$ | 9-9.8 (10.7) | 27-28.5 (30.1) | 12.5-13.5 (14.9) | 9.1-9.7 (10.5) |
| $\gamma$, mJ/mol·K$^2$ | 5.21 | 10.7 | 7.16 | 5.87 |
| $R_\Delta = 2\Delta(0)/k_B T_C$ | 3.74-3.85 (4) | 4.4-4.57 (4.77) | 3.75-3.86 (4.0) | 3.61-3.68 (3.8) |

Estimation of the averaged Fermi velocity $V_F \sim 3.35 \times 10^5$ m/s in hexagonal CeH$_9$ at 120 GPa makes it possible to calculate the London penetration depth $\lambda_L \sim 186$ nm, coherence length $\xi_{BCS} = 34$ nm, and lower critical magnetic field $\mu_0 H_{C1} = 0.013$ T. The critical current density $J_C = en_e V_L$, evaluated by the Landau criterion for superfluidity [32] $V_L = min \frac{\varepsilon(p)}{p} \cong \frac{\Delta_0}{\hbar k_F}$, was calculated to be about $1.5 \times 10^8$ A/cm$^2$.
The calculated Ginzburg–Landau parameter [18] is over 55, which is typical for type II superconductors.

**Table S4.** Additional electronic and superconducting parameters of $P6_3/mmc$-CeH$_9$ at $\mu^* = 0.05$.

| Parameter | 120 GPa | 150 GPa | 200 GPa |
|---|---|---|---|
| $N(E_F)$, states/eV/f.u. | 0.92 | 0.823 | 0.739 |
| Average Fermi velocity $V_F$, m/s | $3.35 \cdot 10^5$ | $2.90 \cdot 10^5$ | $2.60 \cdot 10^5$ |
| London penetration depth $\lambda_L$, nm | 186 | 231 | 272 |
| Coherence length, Å | 34 | 43 | 51 |
| Ginzburg–Landau parameter $\kappa$ | 55 | 54 | 53 |
| Lower critical magnetic field $\mu_0 H_{C1}$, mT | 13.5 | 8.7 | 6.3 |
| Upper critical magnetic field $\mu_0 H_{C2}$, T | 29 | 18 | 12.5 |
| Clogston–Chandrasekhar paramagnetic limit, T | 253 | 173 | 130 |
| Critical current density $J_C$, A/cm$^2$ | $1.47 \cdot 10^8$ | $0.75 \cdot 10^8$ | $0.45 \cdot 10^8$ |

**Table S5.** Partial electron–phonon coupling coefficients $\lambda_q$ of $P6_3/mmc$-CeH$_9$ at 150 GPa on a shifted 4×4×2 $q$-mesh used within the optimized tetrahedron method.

| $q$-point | $q_x$ | $q_y$ | $q_z$ | $\lambda_q$ |
|---|---|---|---|---|
| 1 | 0.126422751 | 0.216464547 | 0.162748325 | 1.200 |
| 2 | 0.128318799 | 0.50548554 | 0.162748325 | 0.774 |
| 3 | 0.122630653 | -0.361577438 | 0.162748325 | 0.816 |
| 4 | 0.124526702 | -0.072556445 | 0.162748325 | 0.789 |
| 5 | 0.377372203 | 0.360372649 | 0.162748325 | 0.776 |
| 6 | 0.379268252 | 0.649393641 | 0.162748325 | 0.968 |
| 7 | 0.373580105 | -0.217669336 | 0.162748325 | 1.442 |
| 8 | 0.375476154 | 0.071351656 | 0.162748325 | 2.980 |

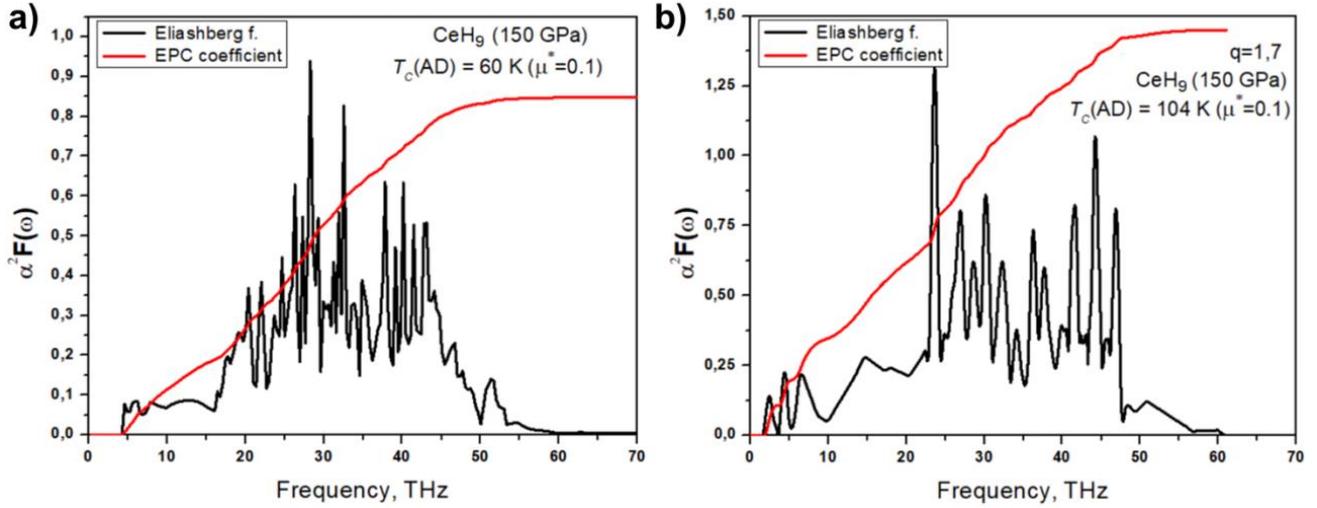

**Fig. S32.** Eliashberg functions calculated for $P6_3/mmc$-CeH$_9$ at 150 GPa within the tetrahedron method: (a) full (averaged over all $q$-points) and (b) partial, average of q$_1$ and q$_7$. "AD" denotes the Allen–Dynes formula[16].

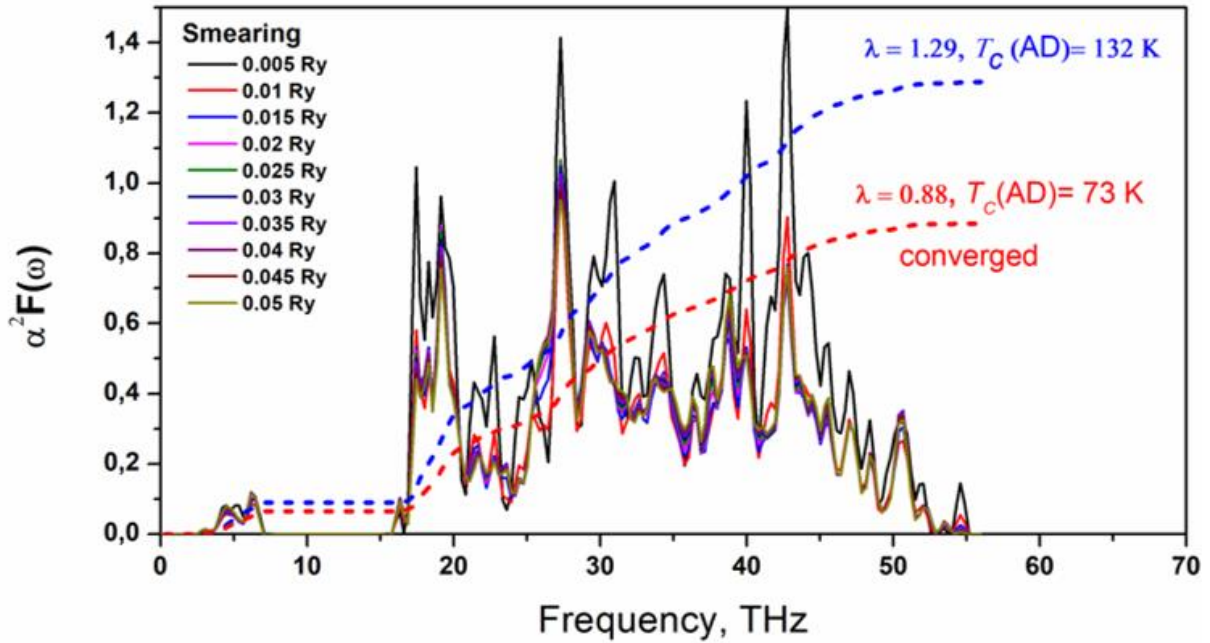

**Fig. S33.** Eliashberg functions for $P6_3/mmc$-CeH$_9$ at 150 GPa with different smearing σ = 0.005–0.05 Ry. The calculations were performed within the interpolation method using the PBE–GGA functional. Converged λ and $T_C$ ~ 73 K (μ* = 0.1) are marked in red, the first α$^2$F (σ = 0.005 Ry) that gives almost 2 times higher $T_C$ is shown with a dashed blue line. "AD" denotes the Allen–Dynes formula[16].

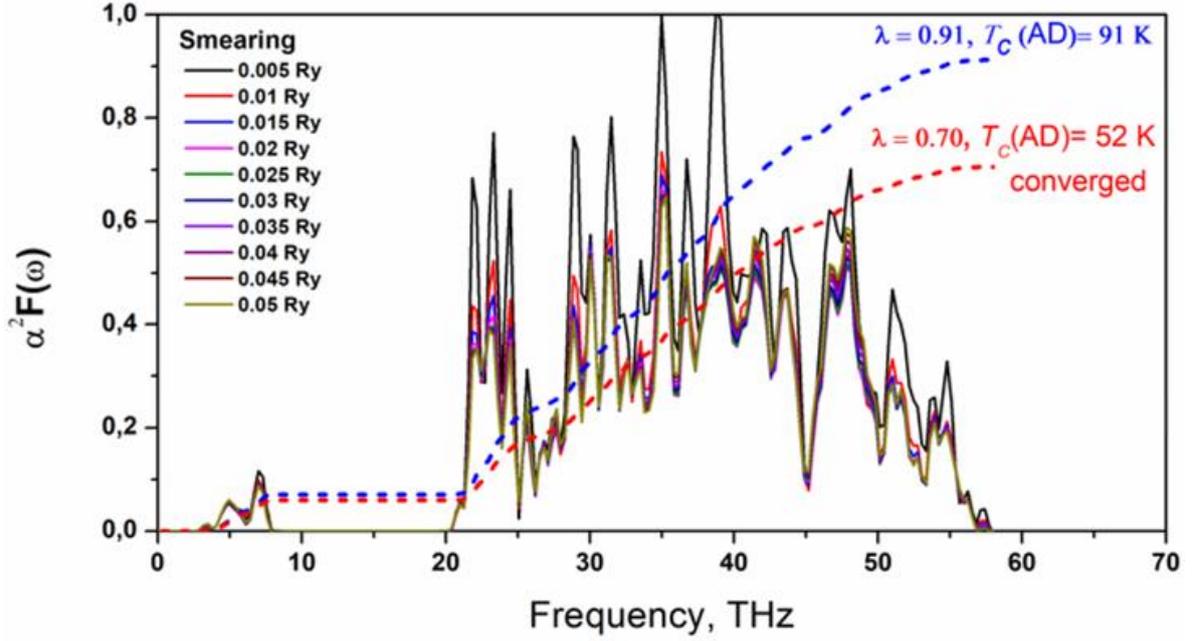

**Fig. S34.** Eliashberg functions for *P6₃/mmc*-CeH₉ at 200 GPa with different smearing σ = 0.005–0.05 Ry. The calculations were performed within the interpolation method using the PBE–GGA functional. Converged λ and $T_C$ ~ 52 K (μ* = 0.1) are marked in red, the first $α^2F$ (σ = 0.005 Ry) that gives almost 2 times higher $T_C$ is shown with a dashed blue line. "AD" denotes the Allen–Dynes formula[16].

**Table S6.** Electron–phonon coupling parameters for *P6₃/mmc*-CeH₉ at 150 GPa calculated with different smearing σ = 0.005–0.05 Ry within the interpolation method using the PBE–GGA functional.

| σ, Ry | λ | λ (from $α^2F$) | $ω_{log}$, K |
|-------|------|---------|---------|
| 0.005 | 1.287445 | 1.287435 | 1239.251 |
| 0.01 | 0.884853 | 0.884846 | 1226.423 |
| 0.015 | 0.885523 | 0.885514 | 1203.938 |
| 0.02 | 0.915781 | 0.915772 | 1195.927 |
| 0.025 | 0.92836 | 0.92835 | 1194.137 |
| 0.03 | 0.932619 | 0.932608 | 1194.483 |
| 0.035 | 0.932256 | 0.932246 | 1195.337 |
| 0.04 | 0.92924 | 0.929229 | 1196.104 |
| 0.045 | 0.924921 | 0.92491 | 1195.9 |
| 0.05 | 0.920389 | 0.920378 | 1194.933 |

**Table S7.** Electron–phonon coupling parameters for *P6₃/mmc*-CeH$_9$ at 200 GPa calculated with different smearing $\sigma = 0.005$–0.05 Ry within the interpolation method using the PBE–GGA functional.

| $\sigma$, Ry | $\lambda$ | $\lambda$ (from $\alpha^2 F$) | $\omega_{log}$, K |
|---|---|---|---|
| 0.005 | 0.912703 | 0.912712 | 1439.232 |
| 0.01 | 0.705513 | 0.705516 | 1425.416 |
| 0.015 | 0.668559 | 0.668559 | 1424.712 |
| 0.02 | 0.650459 | 0.650458 | 1424.41 |
| 0.025 | 0.648303 | 0.648301 | 1424.62 |
| 0.03 | 0.655215 | 0.655213 | 1425.178 |
| 0.035 | 0.663881 | 0.663879 | 1425.716 |
| 0.04 | 0.670909 | 0.670906 | 1425.855 |
| 0.045 | 0.675568 | 0.675565 | 1425.102 |
| 0.05 | 0.678 | 0.677996 | 1423.52 |

# IV. Electronic density of states of CeH₉

**Table S8.** Contributions of different atoms to the total density of states of *P6₃/mmc*-CeH₉ at the Fermi level (in states/eV/Ce).

| Pressure, GPa | H | Ce | Total |
|---|---|---|---|
| 120 | 0.173 | 0.747 | 0.920 |
| 150 | 0.168 | 0.655 | 0.823 |
| 200 | 0.165 | 0.574 | 0.739 |

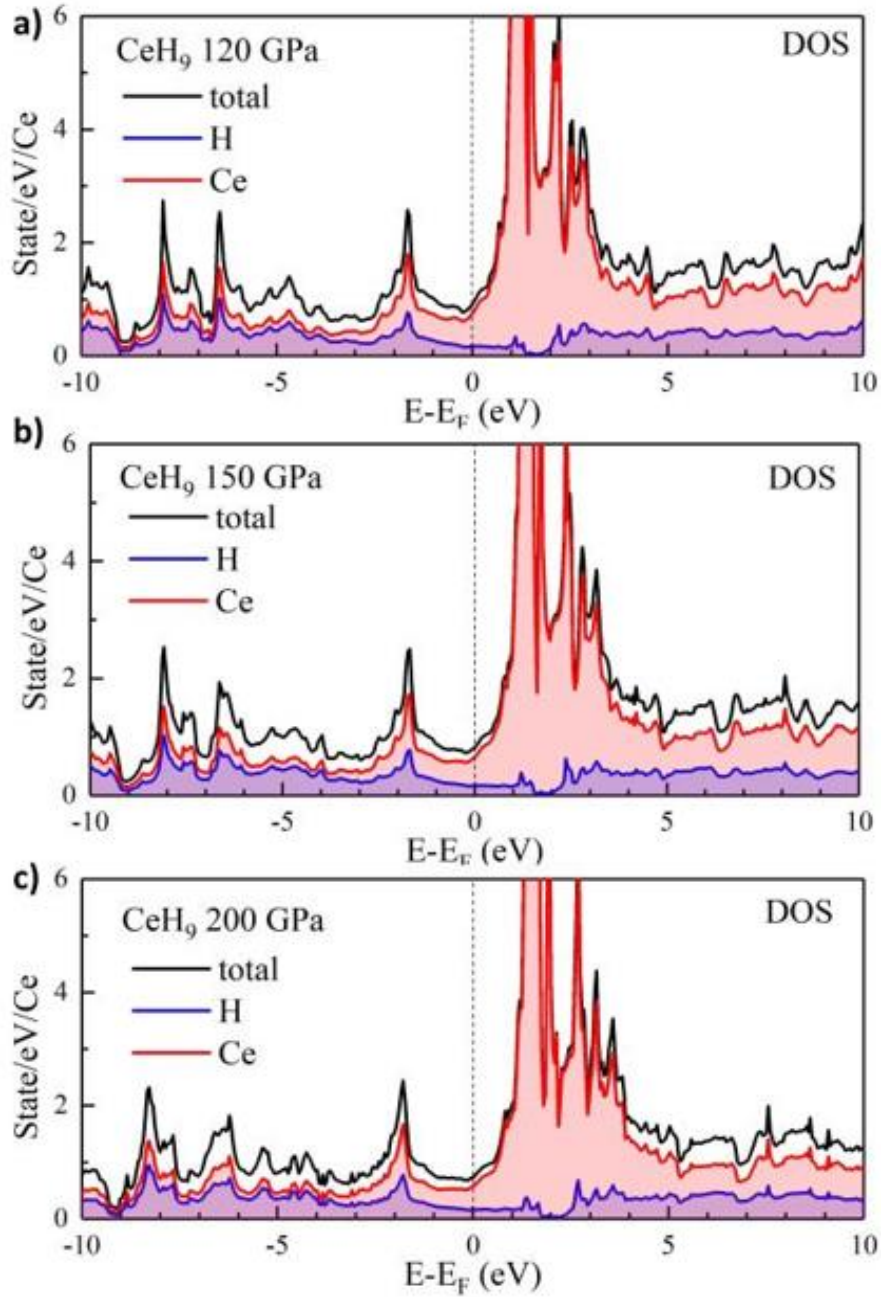

**Fig. S35.** Partial (H and Ce) and total electron density of states of *P6₃/mmc*-CeH₉ at (a) 120 GPa, (b) 150 GPa, and (c) 200 GPa.

# V. Elastic properties

**Table S9.** Elastic properties and thermodynamic parameters of $P6_3/mmc$-CeH$_9$ ($Z = 2$). Calculations with and without the SOC show almost identical results.

| Parameter | 120 GPa | 150 GPa | 150 GPa (SOC) | 200 GPa |
|---|---|---|---|---|
| $a$, Å | 3.653 | 3.591 | 3.591 | 3.507 |
| $c$, Å | 5.485 | 5.364 | 5.363 | 5.198 |
| $V_{DFT}$, Å$^3$ | 63.43 | 59.90 | 59.90 | 55.37 |
| $C_{11}$, GPa | 644 | 747 | 745 | 920 |
| $C_{12}$, GPa | 219 | 242 | 244 | 294 |
| $C_{13}$, GPa | 260 | 304 | 302 | 351 |
| $C_{22}$, GPa | 644 | 747 | 746 | 920 |
| $C_{23}$, GPa | 259 | 304 | 301 | 351 |
| $C_{33}$, GPa | 521 | 584 | 589 | 729 |
| $C_{44}$, GPa | 212 | 252 | 250 | 313 |
| $C_{55}$, GPa | 122 | 191 | 193 | 280 |
| $C_{66}$, GPa | 122 | 191 | 223 | 280 |
| $B$, GPa | 364 | 418 | 418 | 505 |
| $G$, GPa | 158 | 205 | 212 | 276 |
| $E$, GPa | 414 | 529 | 544 | 701 |
| Poisson ratio η | 0.31 | 0.29 | 0.28 | 0.27 |
| Density ρ, kg/m$^3$ | 7801 | 8261 | 8261 | 8937 |
| Transverse sound velocity $v_t$, m/s | 4506 | 4971 | 4971 | 5541 |
| Longitudinal sound velocity $v_l$, m/s | 8586 | 9140 | 9140 | 9871 |
| Debye temperature $\theta_D$, K | 1010 | 1148 | 1167 | 1330 |
| $\omega_{log} = 0.827\theta_D$, K | 835 | 950 | 965 | 1100 |

# References

1   Baer, B. J., Chang, M. E. & Evans, W. J. Raman shift of stressed diamond anvils: Pressure calibration and culet geometry dependence. *J. Appl. Phys.* **104**, 034504 (2008).

2   Parvanov, V. M. *et al.* Materials for hydrogen storage: structure and dynamics of borane ammonia complex. *Dalton Trans.*, 4514-4522 (2008).

3   Prescher, C. & Prakapenka, V. B. DIOPTAS: a program for reduction of two-dimensional X-ray diffraction data and data exploration. *High Press. Res.* **35**, 223-230 (2015).

4   Young, R. A. The Rietveld Method. *Int. Union Cryst.* **30**, 494 (1995).

5   Petříček, V., Dušek, M. & Palatinus, L. Crystallographic Computing System JANA2006: General features. *Z. Kristallogr* **229**, 345-352 (2014).

6   Le Bail, A., Duroy, H. & Fourquet, J. L. Ab-initio structure determination of LiSbWO6 by X-ray powder diffraction. *Mater. Res. Bull.* **23**, 447-452 (1988).

7   Giannozzi, P. *et al.* QUANTUM ESPRESSO: a modular and open-source software project for quantum simulations of materials. *J. Phys. Condens. Matter* **21**, 395502 (2009).

8   Giannozzi, P. *et al.* Advanced capabilities for materials modelling with QUANTUM ESPRESSO. *J. Phys. Condens. Matter* **29**, 465901 (2017).

9   Baroni, S., de Gironcoli, S., Dal Corso, A. & Giannozzi, P. Phonons and related crystal properties from density-functional perturbation theory. *Rev. Mod. Phys.* **73**, 515-562 (2001).

10  Hartwigsen, C., Goedecker, S. & Hutter, J. Relativistic separable dual-space Gaussian pseudopotentials from H to Rn. *Phys. Rev. B* **58**, 3641-3662 (1998).

11  Goedecker, S., Teter, M. & Hutter, J. Separable dual-space Gaussian pseudopotentials. *Phys. Rev. B* **54**, 1703-1710 (1996).

12  Kawamura, M., Gohda, Y. & Tsuneyuki, S. Improved tetrahedron method for the Brillouin-zone integration applicable to response functions. *Phys. Rev. B* **89**, 094515 (2014).

13  Wierzbowska, M. l., Gironcoli, S. d. & Giannozzi, P. Origins of low- and high-pressure discontinuities of Tc in niobium. *Preprint at https://arxiv.org/abs/cond-mat/0504077* (2006).

14  Bergmann, G. & Rainer, D. The sensitivity of the transition temperature to changes in α2F(ω). *Zeitschrift für Physik* **263**, 59-68 (1973).

15  Allen, P. B. & Dynes, R. C. A Computer Program for Numerical Solution of the Eliashberg Equation to Find Tc. *Tech. Rep. 7 TCM41974* (1974).

16  Allen, P. B. & Dynes, R. C. Transition temperature of strong-coupled superconductors reanalyzed. *Phys. Rev. B* **12**, 905-922 (1975).

17  Carbotte, J. P. Properties of boson-exchange superconductors. *Rev. Mod. Phys.* **62**, 1027-1157 (1990).

18  Ginzburg, V. L. & Landau, L. D. On the Theory of superconductivity. *Zh. Eksp. Teor. Fiz.* **20**, 1064-1082 (1950).

19  Ziman, J. M. *Electrons and Phonons.*   (Clarendon Press, Oxford, 1960).

20  Talantsev, E. F. Debye temperature in LaHx-LaDy superconductors. *Preprint at https://arxiv.org/abs/2004.03155* (2020).

21  Hill, R. The Elastic Behaviour of a Crystalline Aggregate. *Proc. Phys. Soc. A* **65**, 349-354 (1952).

22  Anderson, O. L. A simplified method for calculating the debye temperature from elastic constants. *J. Phys. Chem. Solids* **24**, 909-917 (1963).

23  Ravindran, P. *et al.* Density functional theory for calculation of elastic properties of orthorhombic crystals: Application to TiSi2. *J. Appl. Phys.* **84**, 4891-4904 (1998).

24    Li, X. *et al.* Polyhydride CeH9 with an atomic-like hydrogen clathrate structure. *Nat. Commun.* **10**, 3461 (2019).

25    Salke, N. P. *et al.* Synthesis of clathrate cerium superhydride CeH9 at 80-100 GPa with atomic hydrogen sublattice. *Nat. Commun.* **10**, 4453 (2019).

26    Perdew, J. P., Burke, K. & Ernzerhof, M. Generalized Gradient Approximation Made Simple. *Phys. Rev. Lett.* **77**, 3865-3868 (1996).

27    Li, B. *et al.* Predicted high-temperature superconductivity in cerium hydrides at high pressures. *J. Appl. Phys.* **126**, 235901 (2019).

28    Sanna, A. *et al.* Anisotropic gap of superconducting CaC6: A first-principles density functional calculation. *Phys. Rev. B* **75**, 020511 (2007).

29    Yang, T.-R. *et al.* Transport and Magnetic Properties in MgB2. *Int. J. Mod. Phys. B* **17**, 2845-2850 (2003).

30    Gajda, D. *et al.* The critical parameters in in-situ MgB2 wires and tapes with ex-situ MgB2 barrier after hot isostatic pressure, cold drawing, cold rolling and doping. *J. Appl. Phys.* **117**, 173908 (2015).

31    Eliashberg, G. M. Interactions between Electrons and Lattice Vibrations in a Superconductor. *JETP* **11**, 696–702 (1959).

32    Landau, L. Theory of the Superfluidity of Helium II. *Phys. Rev.* **60**, 356-358 (1941).